\documentclass[useAMS,usenatbib]{mn2e}

\input{epsf}

\def\gsim{ \lower .75ex \hbox{$\sim$} \llap{\raise .27ex \hbox{$>$}} }
\def\lsim{ \lower .75ex \hbox{$\sim$} \llap{\raise .27ex \hbox{$<$}} }

\title[Morphological evolution of disk galaxies in clusters]{Morphological evolution of disk galaxies in clusters}

\author[Mastropietro et al.]
{Chiara Mastropietro $^1$ \thanks{E-mail: chiara@physik.unizh.ch}, 
Ben Moore $^1$, Lucio Mayer $^1$, Victor P. Debattista $^{2,3}$, 
\newauthor{Rocco Piffaretti} $^{1,4}$ and Joachim Stadel $^1$\\  
$^1$Institute for Theoretical Physics, University of Z\"urich, CH-8057 Z\"urich, Switzerland\\
$^2$Astronomy Dept., University of Washington Seattle WA 98195-1580\\
$^3$Institut f\"ur Astronomie, ETH Z\"urich, CH-8093 Z\"urich, Switzerland\\
$^4$Institut f\"ur Astrophysik, Leopold-Franzens Universit\"at Innsbruck, A-6020 Innsbruck, Austria
}
\begin{document}


\pagerange{\pageref{firstpage}--\pageref{lastpage}} \pubyear{00} 

\maketitle

\label{firstpage}

\begin{abstract}
The recent discovery of hidden non-axysimmetric and disk-like
structures in bright Virgo dwarf elliptical and lenticular
galaxies (dE/dSph/dS0) indicates that they may have late-type progenitors.
Using N-body simulations we follow the evolution of
disk galaxies within a $\Lambda\textrm{CDM}$ cluster simulated with $10^7$ 
particles, where the hierarchical growth and galaxy harassment are modeled
self-consistently. Most of the galaxies undergo
significant morphological transformation, even at
the outskirts of the cluster, and move
through the Hubble sequence from late type disks to dwarf spheroidals.
None of the disks are completely destroyed therefore they can not be
the progenitors of ultra compact dwarf galaxies (UCDs). 
The time evolution of the simulated galaxies
is compared with unsharp-masked images obtained from VLT data and the 
projected kinematics of our models with the latest high 
resolution spectroscopic studies from the Keck and Palomar telescopes.

\end{abstract}


\section{Introduction}
Early type dwarfs (dE and dSO) are the most common type of galaxies in the
nearby universe, yet their origin and nature are still unknown.
These galaxies are found nearly exclusively near bright galaxies and in groups or clusters. Their continuation or separation from the brighter elliptical sequence is currently debated \citep{Graham, Gavazzi}.
Formation scenarios include mainly two different hypotheses: 1) dEs and dSOs are
primordial galaxies, 2) they are the result of a morphological transformation
of spiral and irregular galaxies accreting into the cluster. This latter idea
is supported by some observations, i.e the relative number density
of dwarf early type galaxies increases with the local galaxy density \citep{Ferguson91}, suggesting
that the environment drives galaxy evolution. \citet{Kormendy} investigated the systematic properties of dark matter halos in late-type and dwarf spheroidal galaxies concluding that they form a single physical sequence as a function of the dark matter core mass. 
\citet{Moore98} have shown that galaxy harassment in
clusters can transform spirals into spheroidals. Harassment can explain the
morphological evolution of the small spiral and irregular galaxies observed in
clusters at redshift $z \sim 0.4$ \citep{Dressler} into dwarf ellipticals in the nearby universe \citep{Moore96}. Moreover the radial velocities of early
type galaxies in the Virgo cluster seem to indicate that these galaxies are
not an old cluster population \citep{Conselice} but originate from infalling field galaxies. 
According to this evolutionary scenario we expect to find nearby cluster
galaxies that are currently undergoing morphological transformation and
retain part of their disk nature. Recent observations confirm that early type
galaxies have a broad range of photometric and kinematical characteristics. 
\citet{Jerjen} and \citet{Barazza02} discovered hidden
spiral structures and bar features in five bright dEs in the Virgo cluster,
concluding that the fraction of early type dwarfs hosting a disk component
could be larger than $20\%$. Spirals and disks were also observed within dwarf
spheroidal galaxies in the Coma 
\citep{Graham} and Fornax clusters \citep{DeRijcke01}.
Furthermore, the degree of rotational
support is found to vary from zero to a value close to one expected for a
galaxy flattened by rotation \citep{DeRijcke01, Simien, Pedraz, Geha02, Geha03, Zee}.\\

The aim of this work is to follow the evolution of  disk
galaxies  orbiting in a cluster environment using high resolution N-body 
simulations and to compare the final
harassed remnants with the latest photometric and spectroscopic data.
Previous studies have suffered from low resolution and they used idealised
cluster models for the initial conditions. We aim to use sufficient resolution
that we can follow the detailed morphological evolution and we evolve the galaxy
models within a cluster selected from a cosmological 
simulation. In addition, the model has
cuspy CDM halos and structural parameters as expected 
in the concordance cosmological model.
This study complements that of \citet{Gnedin} who studied the evolution
of luminous early type disk galaxies within a cosmological context.

The paper is organised as follows. In Section 2 we present the main
characteristics of the galaxy models and the $\Lambda\textrm{CDM}$ cluster used.
The results of the three dimensional simulations are illustrated in Section 3,
while in Section 4 and 5 we report a projected photometric and kinematical
analysis of the simulated remnants and the comparison with observations.

\section{Simulations}
All the simulations have been carried out using PKDGRAV, a
parallel multistepping N--body tree--code designed for high force accuracy \citep{Stadel}.

The galaxy model is a multi--component system with a stellar disk embedded in
a spherical dark matter halo and was constructed using the technique
described by \citet{Hernquist}. The dark matter distribution initially follows
a NFW \citep{Navarro96, Navarro97} profile, which is adiabatically contracted in
response to baryonic infall \citep{Springel}. 
The disk has an exponential surface density profile of the form:
\begin{equation} \label{disk}
\Sigma(R)=\frac{M_d}{2\pi {R_d}^2}\,\textrm{exp}\,(-R/R_d)\, ,
\end{equation}
where $M_d$ and $R_d$ are, respectively, the disk mass and radial 
scale length (in cylindrical coordinates), while the thin vertical 
structure is characterised by the scale height $z_d$ which sets the ``temperature'' of the disk
\begin{equation} \label{diskh} 
\rho_d(R,z) = \frac{\Sigma(R)}{2z_d}\,\textrm{sech}^2\,(z/z_d)\, .
\end{equation}
The structural parameters of the disk and the halo are chosen so that
the resulting rotation curve resembles that of a typical bulgeless
late-type (Sc/Sd) disk galaxy \citep{Courteau, Persic}.
The model parameters are initialised following the same procedure 
as \citet{Mayer02}, which was based on the galaxy formation model of \citet{Mo}.
The mass within the virial radius was set equal to $7 \times 10^{10}$ M$_\odot$   and the
fraction of mass in the disk is $\sim 6\%$. The contribution of the
different components to the global rotation curve, assuming a disk scale
length $R_d= 1.5$ kpc and a concentration $c = 10$ (where $c$ is defined as
$c=r_{vir}/r_{s}$, with $r_{vir}$ and $r_s$ respectively virial and scale
radius of the NFW halo) is plotted in Fig. 1. The halo spin parameter, which sets the disk scale length in our modeling, is $\lambda = 0.045$, where $\lambda$ relates the
angular momentum $J$ and the total energy $E$ of a system with virial mass
$M_{vir}$ through the relation $\lambda=J|E|^{1/2}G^{-1}M_{vir}^{-5/2}$. The model has a central surface brightness $\mu_B = 22 \,\textrm{mag}\,\textrm{arcsec}^{-2}$, assuming an initial $V$-band mass to light ratio $\simeq 4$ GEha et al. 2002 and a $B-V$ color of 0.77 \citep{Zee2}. 
\begin{figure} \label{rotcurve}
\epsfxsize=8truecm
\epsfbox{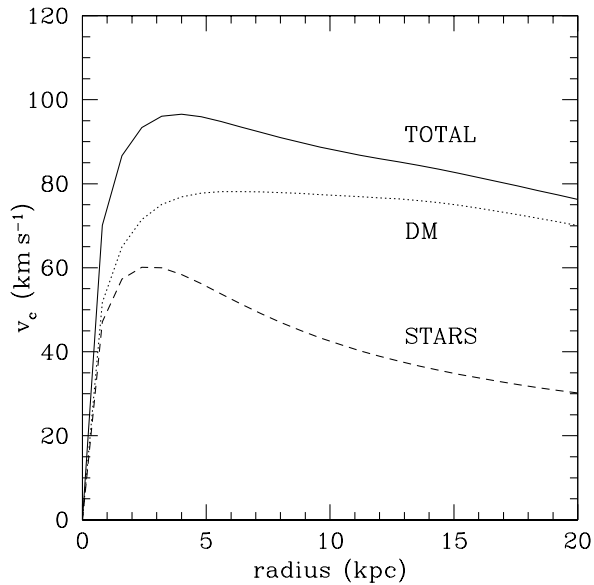}
\caption{The rotation curve of the initial galaxy model. The stellar and dark matter contributions to the total rotation curve are indicated.}
\end{figure}
\begin{figure} \label{qparameter}
\epsfxsize=8truecm
\epsfbox{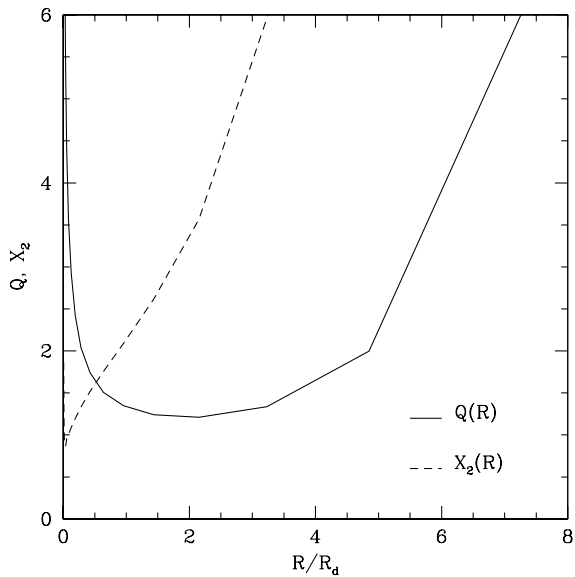}
\caption{$Q$ and $X_2$ parameters as function of radius (in units of the disk scale length $R_d$).}
\end{figure}
The stellar disk satisfies the Toomre \citep{Toomre} stability criterion, which requires, for a rotational supported disk
\begin{equation} \label{Toomres}
Q_{star}(R)= \frac{\sigma_R  k}{3.36 G \Sigma_s}> 1,
\end{equation}
where $\sigma_R$ is the radial velocity dispersion, $\kappa$ is the local epicyclic frequency and $\Sigma_s$ the stellar surface density.
The efficiency of swing amplification of a disk perturbation with m-fold
symmetry is governed by a combination of $Q$ and the parameter $X_m= \kappa ^2 R/(4 \pi m G \Sigma_s)$ \citep{Toomre81}.
The radial profiles of $Q$ and $X_2$ are indicated in Fig. 2. 
\begin{figure*} \label{sequence}
\begin{center}
\vskip 2.0 truein
\hskip -5.2 truein
\includegraphics{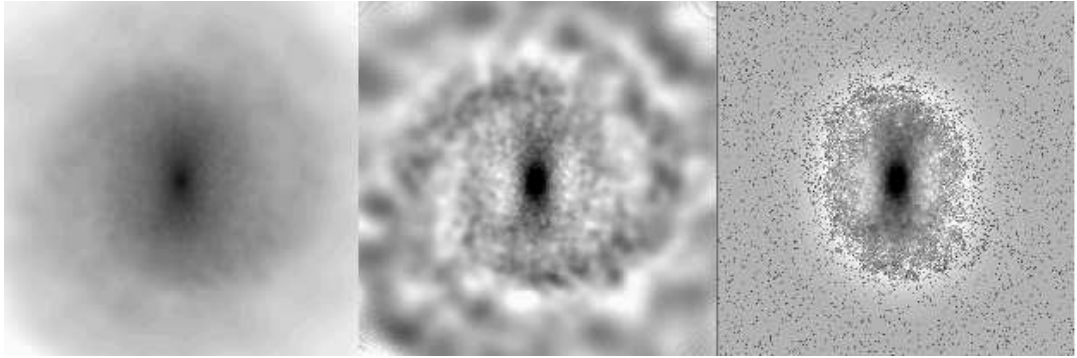}
\vspace{0.7cm}
\caption{Unsharp masking technique applied to the stellar component of a
  simulated face-on galaxy (GAL5 of Table 1). The box size is $\sim 20$ kpc. 
  Left frame: smoothed stellar surface
  density produced from the the simulation output using BEAM by J. Stadel. Middle frame: the result of the unsharp masking on the former image, which reveals presence of a bar and a spiral pattern. Right frame: a noise filter is previously applied to the original surface density map, increasing the intensity of the gray channel for random selected pixels. As a result in the final unsharp image the low density structures disappear and only the bar and a ring feature are still visible. }
\end{center}
\end{figure*}
\begin{figure*} \label{sequence1}
\vskip 4.0 truein
\hskip -2.8 truein
\includegraphics{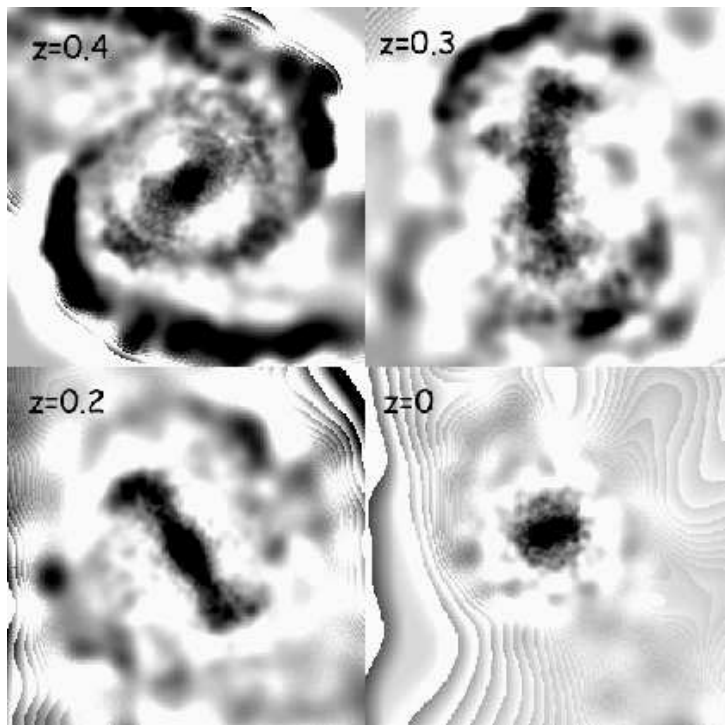}
\vspace{0.5cm}
\caption{Morphological transformation (face-on projection) of a disk galaxy into a dwarf elliptical (GAL8 of Table 1). For each frame the redshift is indicated. The box size is $\sim 20$ kpc. The step like structures visible at the edges of the galaxies are due to features present in the low density regions of the original smoothed stellar surface density maps, which are amplified by the unsharp masking.
}
 \end{figure*}
\begin{figure} \label{buckling}
\epsfxsize=8truecm
\epsfbox{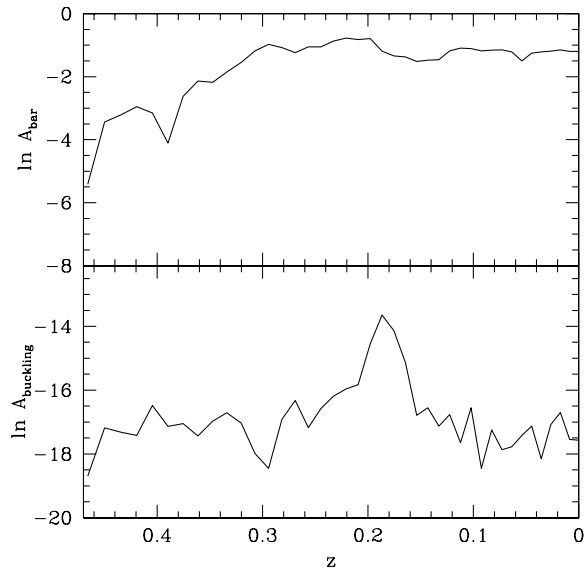}
\caption{Evolution of a representative galaxy (GAL7 of Table 1) from redshift 0.5 to the present time. Top panel: strength of the bar. Bottom: buckling amplitude. The bar forms at $z \sim 0.33$ and strongly buckles at $z \sim 0.2$.    }
\end{figure}
\begin{figure} \label{buckling2}
\epsfxsize=8truecm
\epsfbox{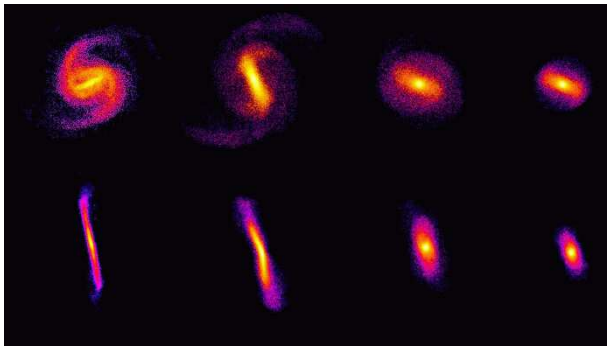}
\caption{Evolution of the stellar surface density of GAL7 from $z=0.3$ (left panels) to $ z\sim 0$ (right panels). The color scale is logarithmic. The top panels represent the face on projection while the corresponding edge on projection is shown on the bottom. The buckling phase, which has a peak at $z=0.2$, does not destroy the bar but makes the central vertical structure of the galaxy thicker.}
\end{figure}

Each galaxy is modeled with $10^6$ particles, $10^5$ of which are in the
disk. The gravitational softening is set to 0.5 kpc for the dark halo and 0.1 kpc for the disk particles.  
The galaxies were evolved for $\sim$ 8.4 Gyrs in a $\Lambda$CDM cluster with
a mass at $z=0$ comparable to that of Virgo ($M_{vir} = 3.1 \times 10^{14}$ M$_\odot$) and a
virial radius $R_{vir}$ of 1.8 Mpc. The cluster, selected from a 300 Mpc cube
simulation \citep{Diemand04b} with cosmological parameters $\Omega_{\Lambda} = 0.732$,
$\Omega_{m} = 0.268$ and $\sigma_8 = 0.9$ \citep{Spergel}, was resimulated at
high resolution and corresponds to the intermediate
resolution cluster D9 of \citet{Diemand04a}. 
The number of particles within the virial radius is $\sim 6 \times 10^6$
and the softening length is 2.4 kpc.

At $z = 0.5$ we replaced 20 random cluster particles (10 within the virial
radius and 10 within 20\% of the virial radius) with the high resolution
galaxy model. Only one galaxy model is used, in order to isolate the effects of orbit on the evolution.  
We assigned to the center of mass of each galaxy the same
orbital position and velocity of the replaced particle. 
We replace particles rather than halos so as not to bias the results by omitting ``overmerged'' halos.
Internal velocities
and lengths were rescaled according to the redshift and cosmology.
In particular the comoving velocities at a given redshift $z$ are expressed in
terms of the Hubble constant $H(z)$ through the relation
\begin{equation} \label{vcomoving}
v = \frac{\dot{r}}{a} - H(z)\,x,
\end{equation}
where $r$ and $x$ are the physical and comoving coordinate and
$a$ the scale factor $1/(1+z)$, respectively.

\section{Galaxy evolution}

Among the initial sample of 20 galaxies, only 13 are still within the
cluster virial radius at $z=0$, while the others lie on bound orbits between
$R_{vir}$ and 1.5 $R_{vir}$. This result is consistent with the fact that
about half of the halos which are presently at the outskirts of the cluster
(up to 2$R_{vir}$) have a pericenter smaller than $R_{vir}$ \citep{Moore04}.

The evolution of many of the galaxies within the cluster is quite violent and due to a sequence of strong gravitational 
encounters with substructures and with the global cluster potential.
Most of the central galaxies lose a significant fraction of
stars and undergo a complete morphological transformation from disks to
spheroidal systems.
In order to highlight the presence of hidden structures we applied
to the stellar component of the simulation results the unsharp masking technique described in Barazza et al. 2002. This method consists of the following steps.
The intensity of each pixel is replaced with the mean intensity of a given area around the pixel producing a smoothed image. The linear dimension of this area
has to be close to the typical scale of the features that we want to
uncover. The original image is then divided by the smoothed one.
Fig. 3 illustrates the results of the unsharp masking applied to a
galaxy of our sample, where the ESO's MIDAS task FILTER/SMOOTH is used for the image smoothing with a smoothing length of 30-40 pixels, corresponding to $\sim 2$ kpc. 

\begin{figure*} \label{tris}
\vskip 2.0 truein
\hskip -5.5 truein
\includegraphics{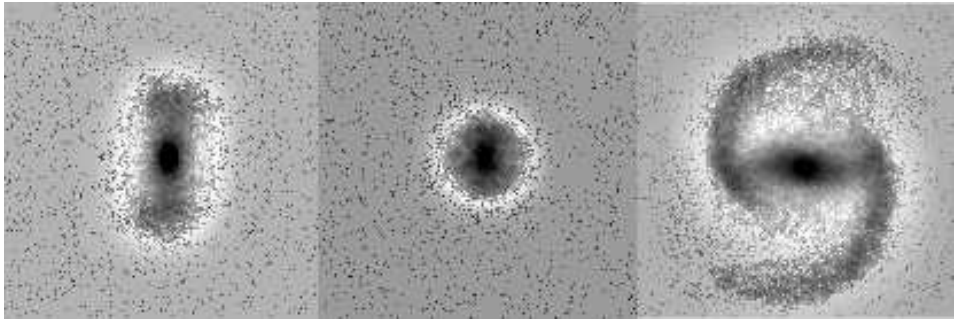}
\vspace{0.7cm}
\caption{Three different final states of the harassed galaxies (unsharp nosified images). From the left
  to the right: bar-like galaxy (GAL7 of Table 1), spherical spheroid (GAL6) and spiral galaxy (GAL3).  The
  addition of noises often hides ring structures surrounding the central bar,
  which is the case of the first remnant on the left. 
 In order to point out the characteristic strucure of the remnants we used two physical scales: the two images on the left have a box of $\sim 12$ kpc while the spiral galaxy on the right has a diameter of $\sim 20$ kpc.
}
\end{figure*}

Fig. 4 represents the different phases of the evolution of a
galaxy  transforming into a dE. 
The first stage of the evolution is characterised by the formation of a strong
bar \citep{Moore98} and an open spiral pattern in the disk. 
The spiral arms are easily stripped while the remaining material forms a
ring structure around the bar. The bar+ring phase is quite stable and is 
apparent in the final state of several of the remnant galaxies. Often the
ring is not directly visible in the simulation output, but is clearly seen after applying the
unsharp masking. When gravitational heating due to tidal interactions removes the remaining spiral
and ring features, only a naked bar remains. The bar usually undergoes a strong buckling instability
and as a result the central part of the galaxy becomes more spherical. 
The buckled bar is subjected to further tidal heating and loses 
mass from its edges, becoming rounder the more encounters it suffers and the longer it evolves near to the cluster centre.
Fig. 5 illustrates the evolution of bar and buckling instabilities for a
representative galaxy which is orbiting close to the center of the
cluster at $z=0$. $A_{bar}$ and $A_{buckling}$ are the bar and buckling amplitudes, defined respectively as the Fourier decomposition of the phase of particles in the xy plane and of their vertical positions \citep{Sparke}
\begin{equation}
A_{\rm bar} = \frac{1}{N} \left| \sum_j e^{2 i \phi_j} \right|
\end{equation}
and
\begin{equation}
A_{\rm buckling} = \frac{1}{N} \left| \sum_j z_j e^{2 i \phi_j} \right|.
\end{equation}
\begin{figure} \label{stream}
\epsfxsize=8truecm
\epsfbox{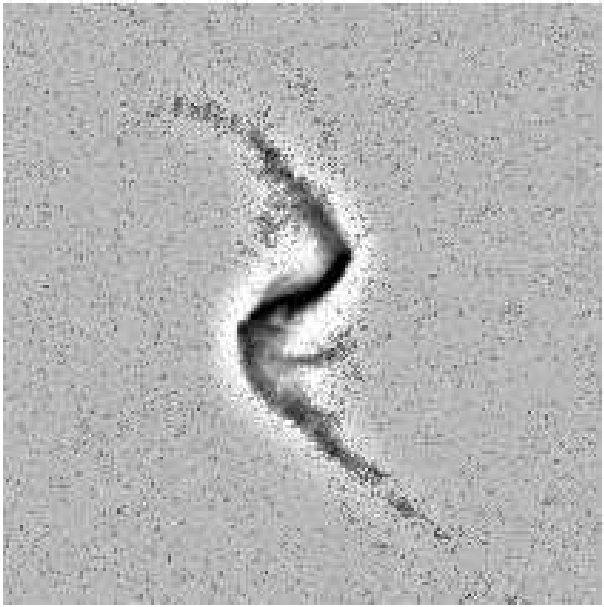}
\caption{Tidal tails of GAL6 (Table 1) generated during a pericentric passage (unsharp nosified image). These smooth features are visible only for a short time (typically about 0.2-0.3 Gyr) since they
are heated by continued encounters with cluster substructures.}
\end{figure}
\begin{figure} \label{icl}
\epsfxsize=8truecm
\epsfbox{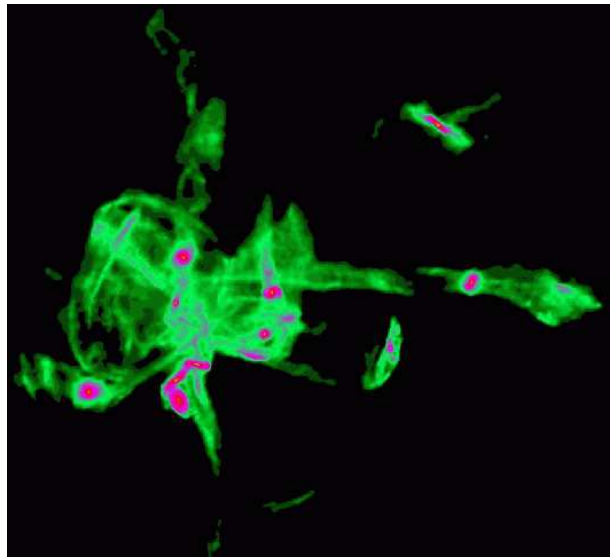}
\caption{Intracluster light produced by tidal processes. The color contrast is chosen in such a way as to point out the low density regions (light green, corresponding to a surface brightness $\mu \lsim 30$ mag arcsec$^{-2}$) and the traces of stellar streams.}
\end{figure}
\begin{figure} \label{c/a}
\epsfxsize=8truecm
\epsfbox{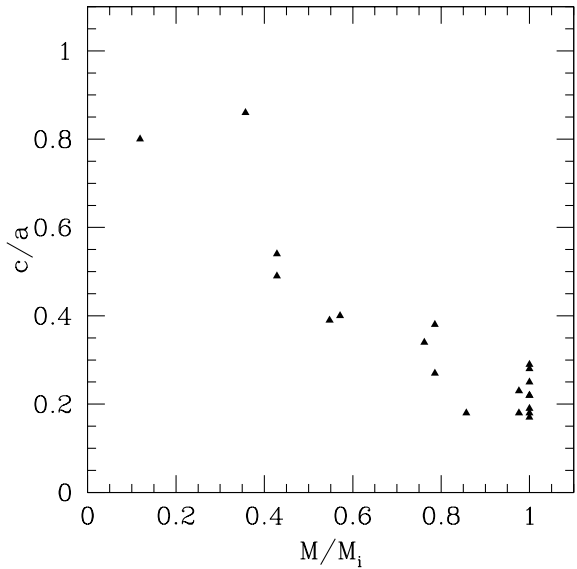}
\caption{Final flattening of the remnants plotted versus the fraction of
  stellar mass within the tidal radius at $z=0$.}
\end{figure}
Fig. 6 shows a face and edge on projection of the same galaxy during
the buckling phase where the bar appeared at $z = 0.3$.  

Depending on the orbit and on the number of close encounters
that the galaxy suffers, the morphological transformation of the stellar
component can be complete and produce a pure spheroidal system, or can lead to
the formation of a more elongated structure, which still retains disk and bar
features (Fig. 7). For a few galaxies orbiting closest to the cluster
center the tidal effects from the
cluster potential become dominant and the evolution is very rapid. Fig. 8
shows the tidal tails generated by a strong impulsive encounter. The
stripped mass is a large fraction of the total stellar mass and the
surface density in the tails is comparable with that of the outer disk. We do not expect that
this kind of structures will be easily observable in clusters, since their life
time is quit short ($\approx 0.2$ Gyrs) due to numerous encounters with other
substructures which heat and dissipate the stripped material.         
Stripped stars tend to create a diffuse distribution around the remnants (see Fig. 9) and in regions close to the center 
it is almost impossible to distinguish well defined stellar streams, even from the debris of a small fraction of the true
number of dwarf galaxies in clusters.

\begin{table} 
  \begin{center}
\begin{tabular}{l|c|c|c|c|c|c|c}
\hline
  Galaxy&Apo-Peri& $r_t$  &$r_e$ &$M_h$ &$M_s$ & $b/a$ & $c/a$  \\

 \hline
  GAL1&200-90   & 8.5 & 2.0 & 2.6 & 2.3 & 0.79 & 0.39\\
  GAL2&150-100  & 10.1 & 1.3 & 2.8 & 2.4 & 0.57 & 0.40 \\ 
  GAL3&650-60   & 16.6 & 2.7 & 13.0 & 4.2 & 0.42 & 0.18 \\
  GAL4&300-50   & 11.6 & 1.8 & 5.2 & 3.2 & 0.44 & 0.34 \\
  GAL5&630-100  & 19.4 & 1.9 & 7.6 & 3.3 & 0.60 & 0.27 \\
  GAL6&90-40    & 3.6 & 0.7 & 0.2 & 0.5 & 0.81 & 0.80 \\
  GAL7&75-210    & 11.0 & 1.8 & 4.2 & 3.3 & 0.61 & 0.38 \\
  GAL8&100-60   & 4.6 & 1.1 & 1.0 & 1.5 & 0.87 & 0.86 \\
  GAL9&120-70   & 7.9 & 1.1 & 1.5 & 1.8 & 0.66 & 0.54\\
  GAL10&270-80   & 9.0 & 1.2 & 1.8 & 1.8 & 0.62 & 0.49 \\
  GAL11&...-460  & 17.2 & 2.0 & 8.2 & 3.6 & 0.71 & 0.18 \\
  GAL12&1130-120 & 16.9 & 2.6 & 14.6 & 4.1 & 0.66 & 0.18 \\
  GAL13&...-900  & 17.4 & 2.6 & 19.2 & 4.2 & 0.41 & 0.28 \\ 
  GAL14&...-...  & 16.3 & 2.5 & 24.4 & 4.2 & 0.79 & 0.29 \\ 
  GAL15&2480-... & 17.0 & 2.9 & 21.1 & 4.2 & 0.43 & 0.19 \\ 
  GAL16&720-210  & 14.7 & 2.6 & 18.8 & 4.1 & 0.45 & 0.23 \\
  GAL17&...-...  & 15.6 & 2.6 & 21.0 & 4.2 & 0.38 & 0.22 \\
  GAL18&2290-... & 15.2 & 2.7 & 19.5 & 4.2 & 0.41 & 0.25 \\
  GAL19&2810-... & 15.0 & 3.0 & 19.6 & 4.2 & 0.77 & 0.17 \\
  GAL20&1720-450 & 16.8 & 2.9 & 21.6 & 4.2 & 0.34 & 0.22 \\ 
\hline
\end{tabular}
 \caption{Final state of the remnants. The second column indicates, when
 available, the apocentric and pericentric distances of the last orbit. 
 For each  galaxy we list three dimensional (spherically averaged) tidal $r_t$ and effective 
$r_e$ radius (in kpc),
 dark matter $M_h$ and stellar $M_s$ mass (in units of $10^9 M_{\odot}$). The
 tidal radius is calculated using SKID \citep{Stadel}, while $r_e$ is the
 radius containing half of the light. The last two columns indicate the axial ratios (respectively intermediate and short axis to major axis)
 measured within 2$r_e$.}
  \end{center}
\end{table}

Table 1 summarises the main properties of the remnant galaxies. 
The initial model has a virial radius of 85 kpc and the effective radius of
the stellar distribution is 2.5 kpc, defined as the radius containing half of the light
of the galaxy.
While almost all the galaxies lose more than two thirds of the dark matter
halo, galaxies located at outskirts of the cluster do not experience a
significative amount of star loss, even if their stellar structure can be
perturbed producing open spiral patterns
and asymmetric features (last frame on the right of Fig. 7). For these remnants the reduction of the
effective three--dimensional radius $r_e$ is related to the formation of a bar that increases the
central stellar density. On the other hand, galaxies orbiting closer to the
cluster center lose up to 90\% of their stellar mass and the decrease of $r_e$
corresponds to a real decrease in the size of the stellar
component. In a few cases the galaxy loses much more dark matter than stars and becomes baryon dominated in the central region. This is due to the fact that the orbits of the dark matter particles in the halo are more 
eccentric than those of the star particles
and therefore reach the tidal radius more easily.
Table 1 shows that the loss of stellar mass
is not a simple function of the mean orbital radius or apocentric/pericentric 
distance from the cluster center, confirming the
importance of harassment. 
The last two columns of Table 1 give the mean axial ratios within
2$r_e$. There is a clear correlation between the c/a ratio (the flattening of the stellar component) and the final stellar mass. As shown in Fig. 10, massive stellar remnants have smaller c/a values and retain more 
of their initial disk nature, while galaxies that have lost most of their 
stars tend to be more spheroidal. 
The final state is always quite prolate, with the exception of GAL6 and GAL8 (the first two points in the upper left corner of Fig. 10)  which are close to spherical. In the case of the less perturbed 
galaxies this is due to the fact that the radius 2$r_e$ defines an area 
roughly corresponding to the bar region, while most of the other galaxies retain part 
of the radial anisotropy that originated during the bar phase \citep{Mayer01}.
\section{Photometric analysis}
In order to compare the results of our simulations with observations and
in particular with \citet{Barazza03}, we selected the remnants within the virial
radius at $z=0$ (first 13 galaxies of Table 1) 
and projected them along a random line of sight.
\begin{figure} \label{Gal9}
\epsfxsize=5truecm
 \hskip 0.5 truein
\epsfbox{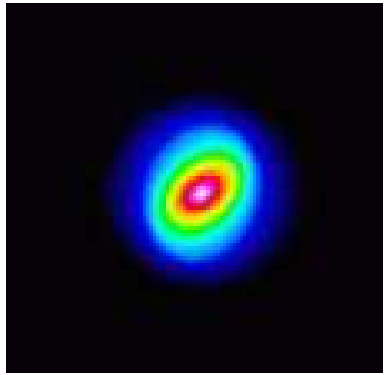}
\caption{Elliptical contours of GAL9. A strong isophotal twist is evident in the outer region (see also Fig. 12).}
\end{figure}
 
\begin{figure*} \label{ph1}
\vskip 5.1 truein
\hskip -4.5 truein
\includegraphics{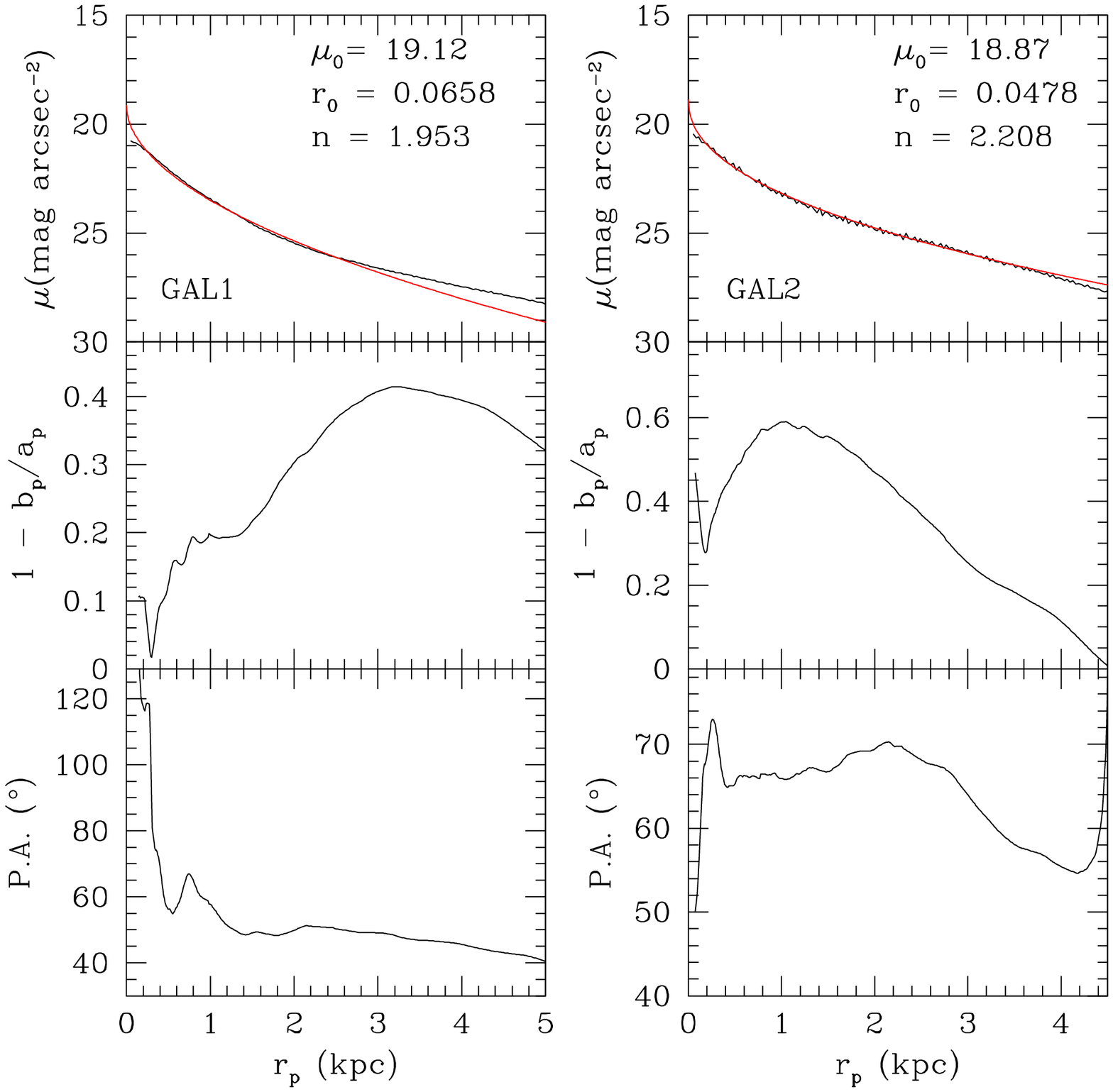}
\vspace{-2.0cm}
\end{figure*}
\begin{figure*} \label{ph2}
\vskip 4.7 truein
\hskip -4.5 truein
\includegraphics{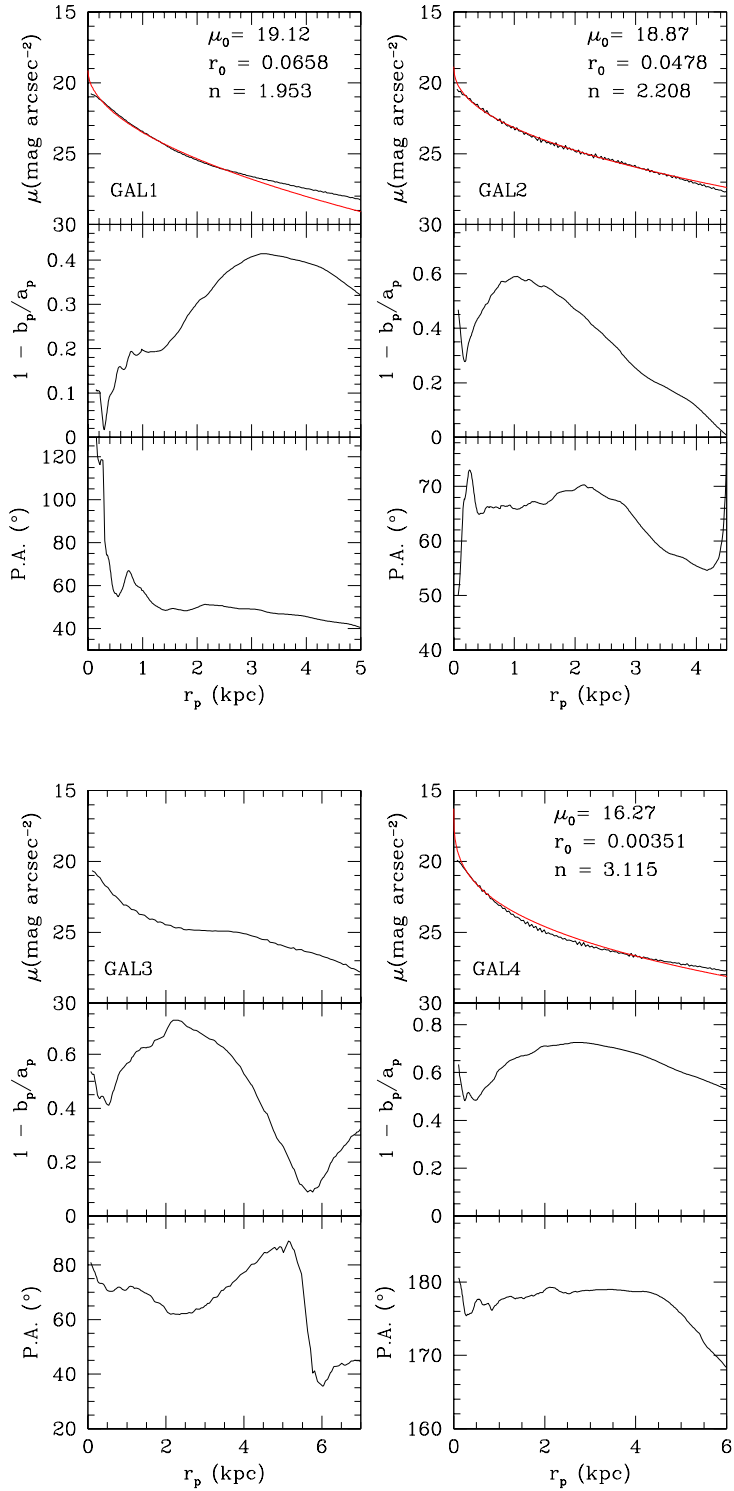}
\vspace{-2.0cm}
\caption{Each set of three vertical panels represents the photometric characteristics of a galaxy from Table 1: $B$-band surface brightness (top panel), 
  ellipticity
  (center) and position angle profile (bottom).  The coordinate $r_p$ is the
  equivalent radius $r_p = \sqrt{a_pb_p}$, where $a_p$ and $b_p$ are the major and minor axis of the projected remnant. The S\'ersic model is superimposed on the surface brightness
  profile (red solid line) and the corresponding best fit
  parameters are given in the top right of the panel. The central surface
  brightness $\mu_0$ is expressed in $\textrm{mag}\, \textrm{arcsec}^{-2}$,
  $r_0$ is in kpc and the shape parameter $n$ is a pure number which
  represents the deviation of the profile from an exponential law.  }
\end{figure*}
\addtocounter{figure}{-1}
\begin{figure*} \label{ph3}
\vskip 5.1 truein
\hskip -4.5 truein
\includegraphics{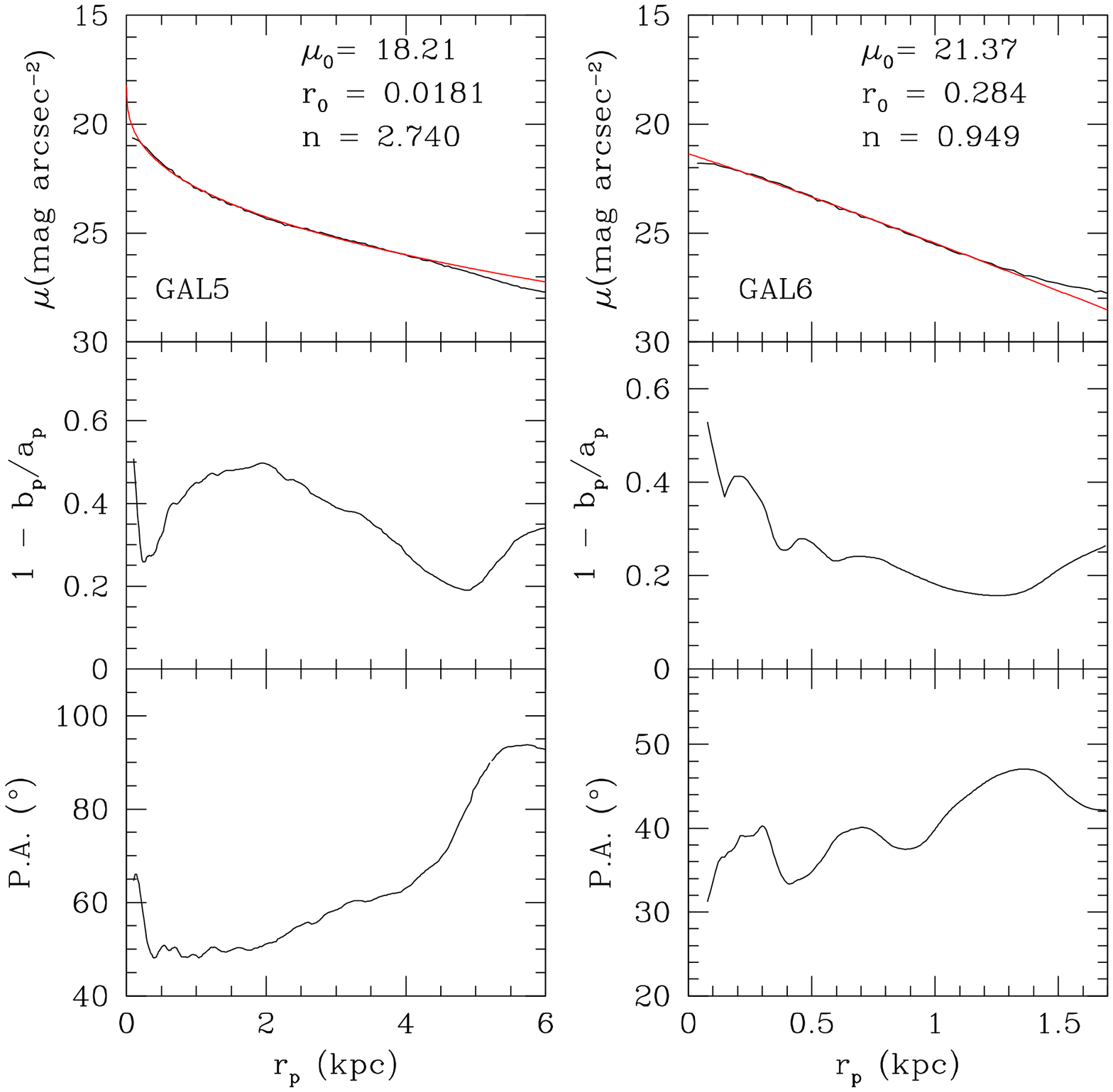}
\vspace{-2.5cm}
\end{figure*}
\begin{figure*} \label{ph4}
\vskip 5.1 truein
\hskip -4.5 truein
\includegraphics{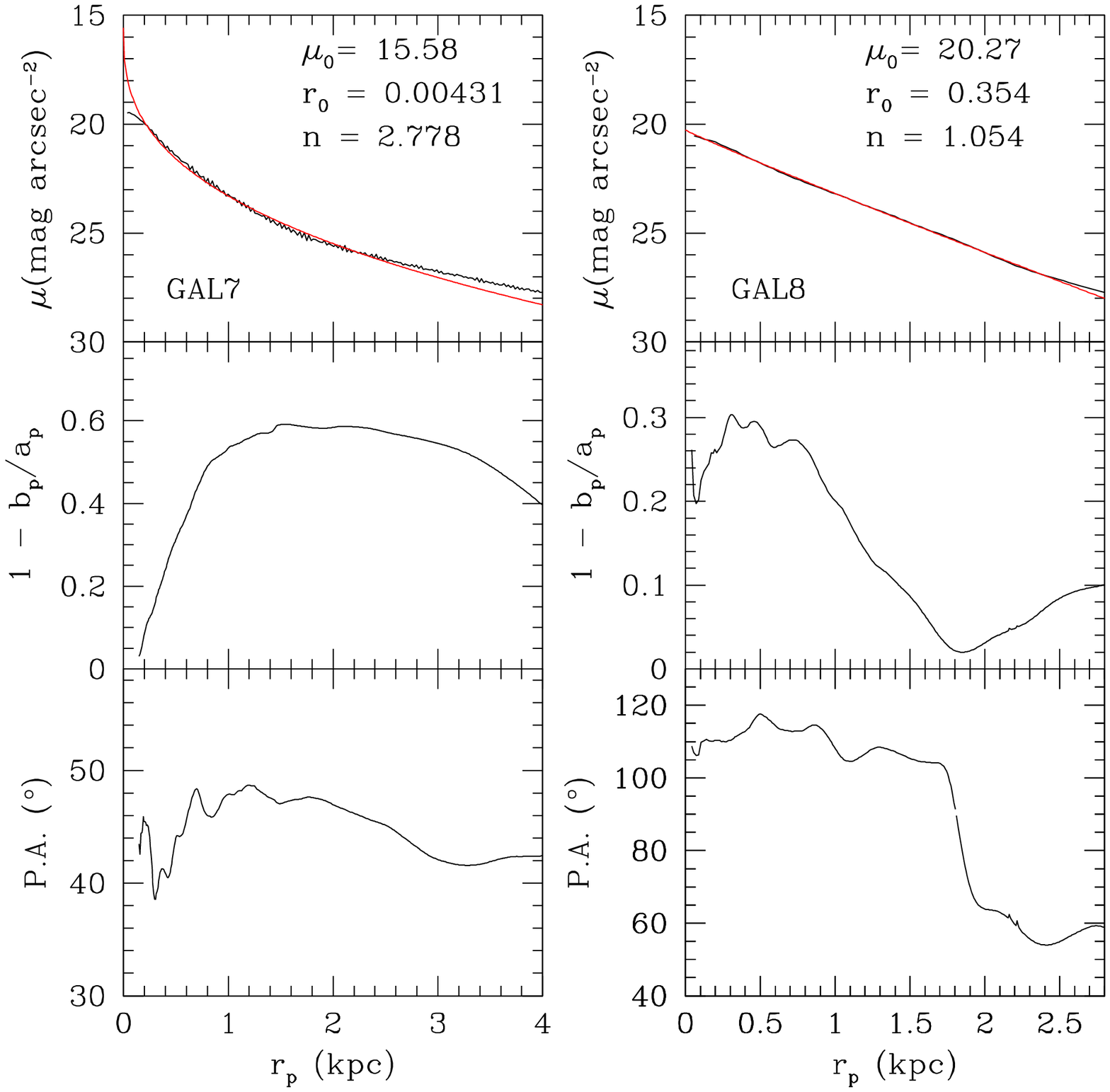}
\vspace{-2.5cm}
\caption{continued.}
\end{figure*}

\addtocounter{figure}{-1}
\begin{figure*} \label{ph5}
\vskip 5.1 truein
\hskip -4.5 truein
\includegraphics{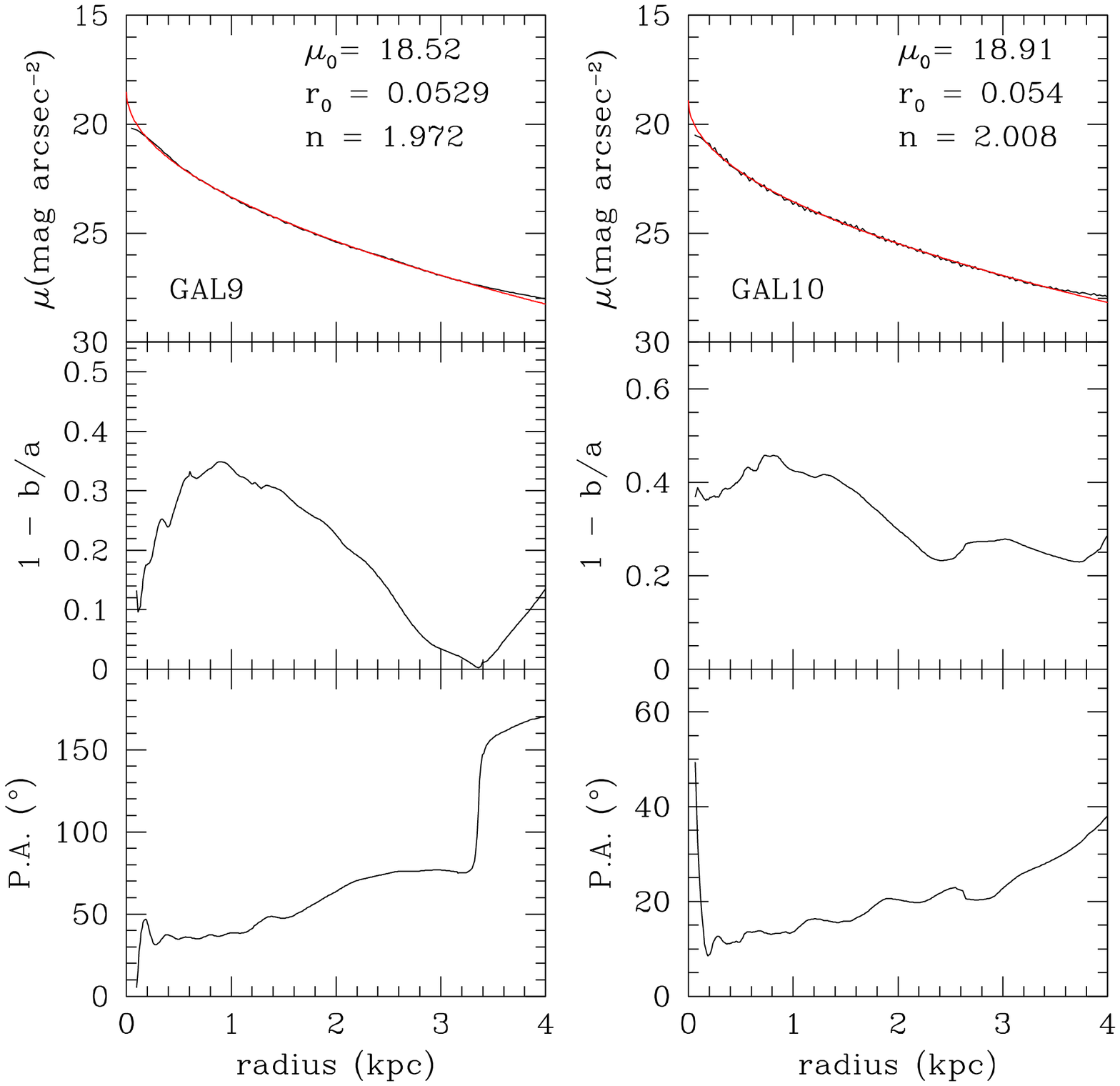}
\vspace{-2.5cm}
\end{figure*}
\begin{figure*} \label{ph6}
\vskip 5.1 truein
\hskip -4.5 truein
\includegraphics{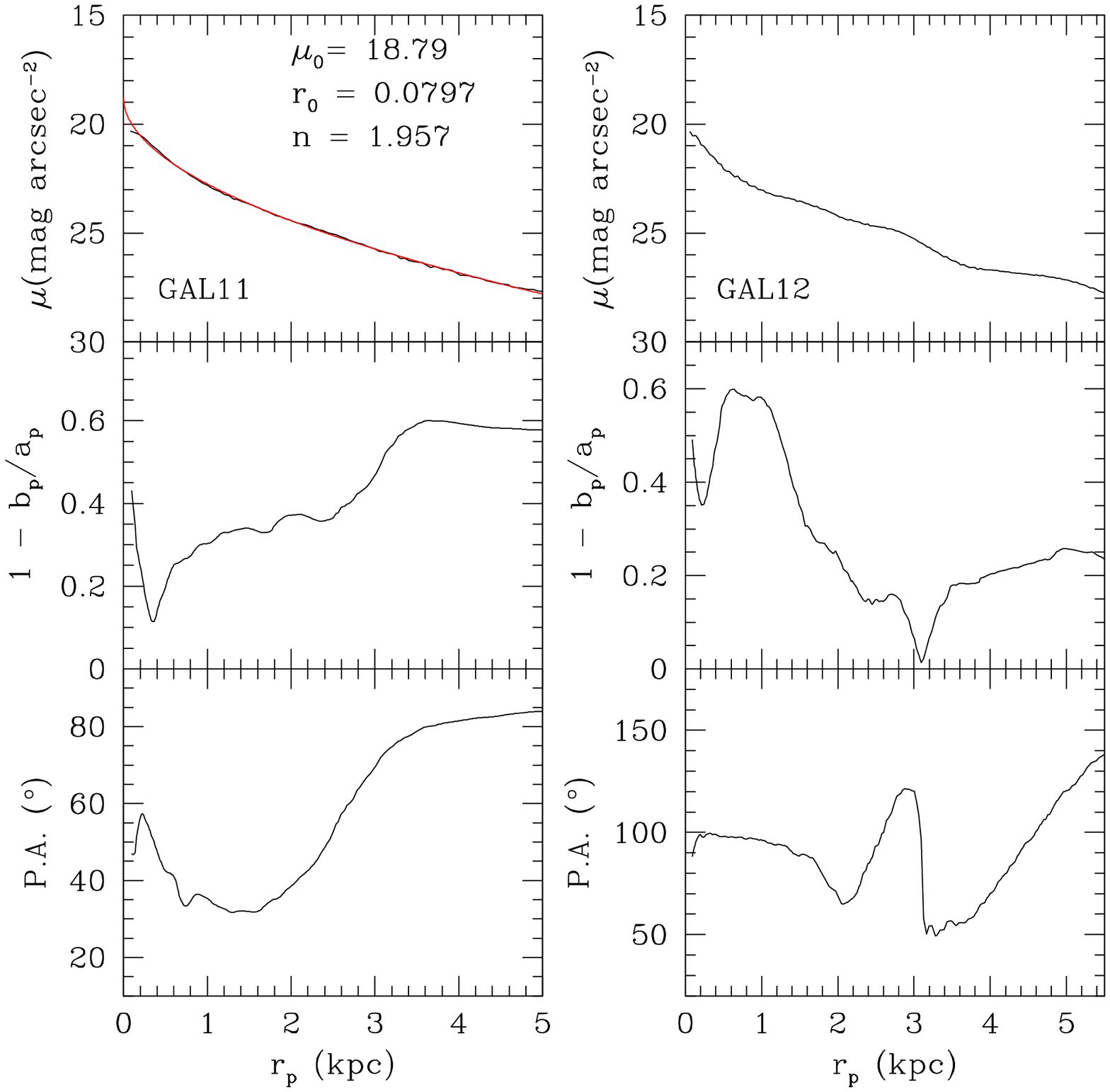}
\vspace{-2.5cm}
\caption{continued.}
\end{figure*}

\addtocounter{figure}{-1}
\begin{figure*} \label{ph7}
\vskip 5.1 truein
\hskip -4.5 truein
\includegraphics{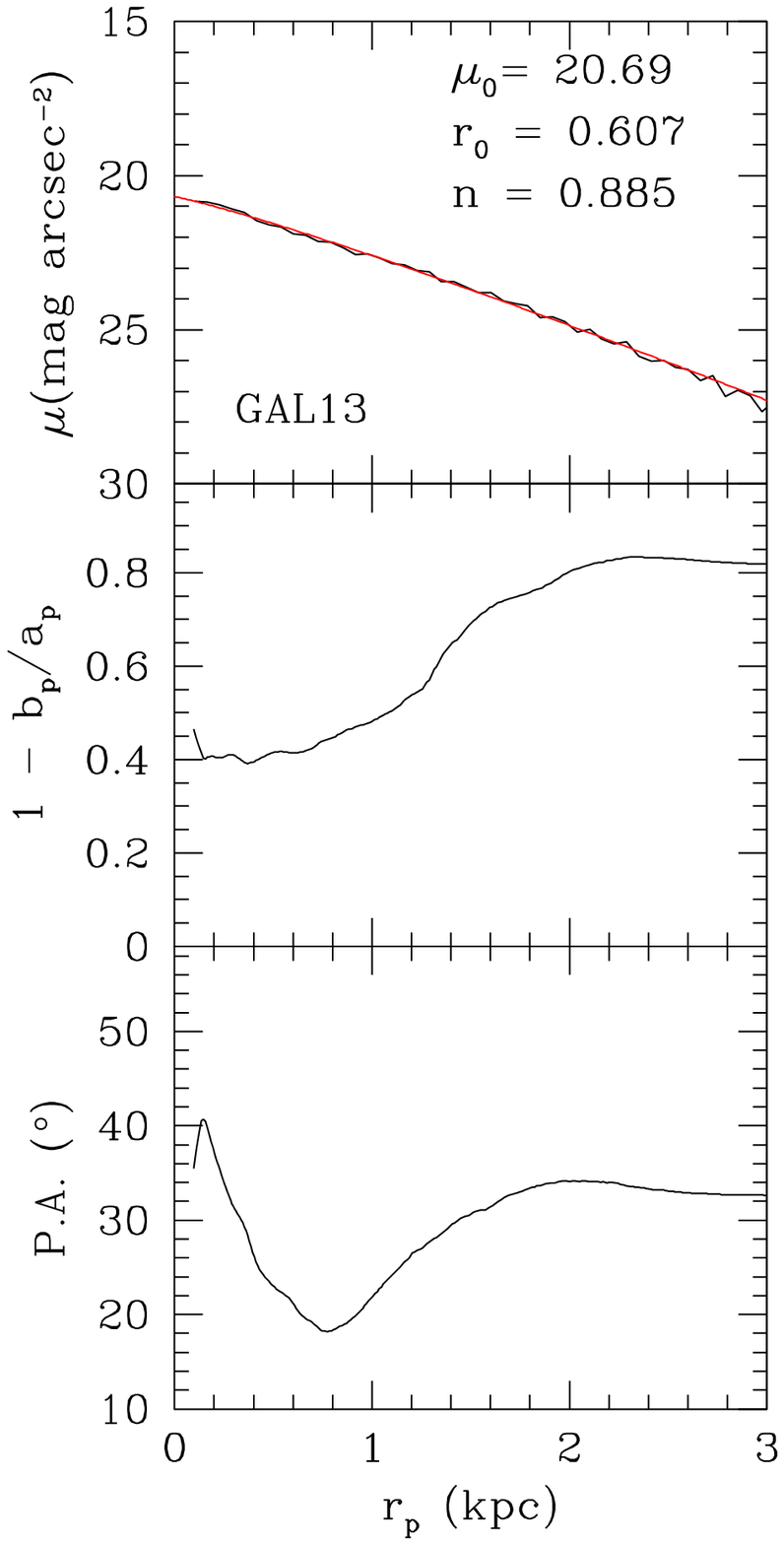}
\vspace{-2.5cm}
\caption{continued.}
\end{figure*}

\begin{figure*} \label{ph7bis}
\vskip 5.1 truein
\hskip -4.5 truein
\includegraphics{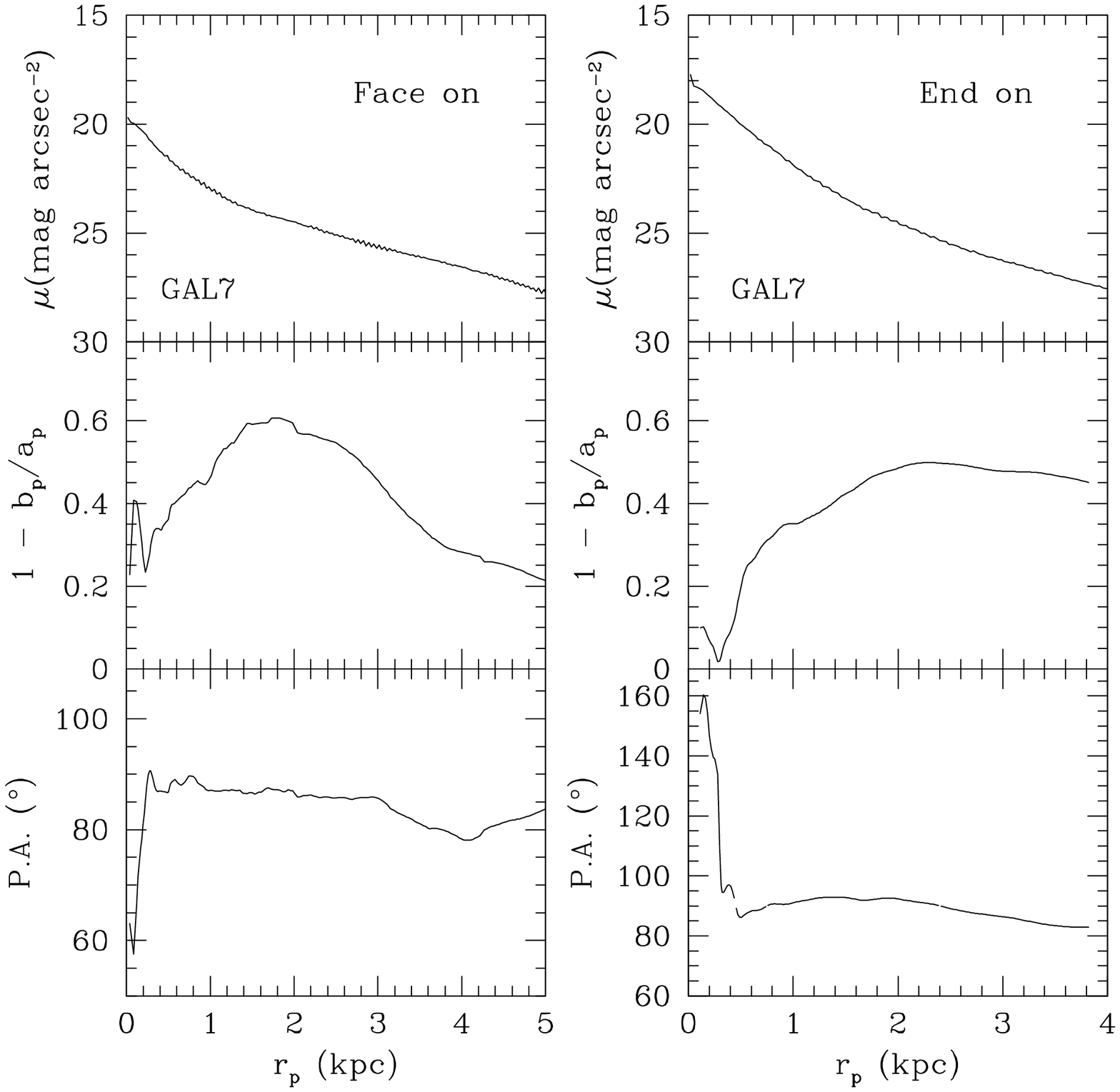}
\vspace{-2.5cm}
\caption{$B$-band surface brightness, ellipticity and position angle profiles of the face on and end on projections of GAL7. }
\end{figure*}

The isodensity contours of the projected disks were fitted with ellipses using the MIDAS task FIT/ELL3. The ellipses centers, ellipticity and position angle were considered as free parameters of the fitting procedure.
The isophotal contours of a representative galaxy (GAL9) are shown in Fig 11. Ellipticity and position angle profiles are plotted in Fig. 12, where the independent variable is the equivalent radius $r_p =
\sqrt{a_pb_p}$, and $a_p$ and $b_p$ are the major and minor axis of the projected remnant.
Even though there is not a well defined trend in the ellipticity profiles, galaxies which
preserve a bar component are generally characterised by large ellipticities in
the central 2-3 kpc and by a drop in the outer regions. Often the inner 0.5 kpc
exhibits a very low ellipticity corresponding to a small nuclear component
produced by buckling instabilities, whereas the external drop is more likely
due to disk heating. Rounder systems have a wider range of profiles,
with eccentricities that may even increase with radius, as in the case of GAL1.

The next step of this analysis consists in calculating the surface brightness profiles of the projected remnants.
In order to derive a luminosity we need to assign a mass to light ratio.
We expect that the ISM would be rapidly stripped by ram pressure effects
such that star formation is truncated shortly after the galaxies enter
the cluster environment. In this case, roughly 8 Gyrs from the beginning of the simulation, 
the stellar mass to light ratio will have increased by about a factor of $1.5-2$
\citep{Mayer01}. For this reason we used a $B$-band $M/L \sim 6$ for the final remnants.
The cumulative light profiles were determined by integrating the light of each
galaxy in elliptical apertures with increasing equivalent radius (MIDAS task
INTEGRATE/ELLIPS) and fixed center, ellipticity and position angle. For these parameters we adopted their mean values within a radius corresponding to half of
the disk size, in order to avoid uncertainties in the isophote fitting procedure due to the low density external regions.  
In Table 2 we list the effective radius of each projected galaxy defined by $r_{ep}$ which is the radius at which 
the cumulative light profile reaches half of the value that it
has at the isophotal level of $28\, \textrm{mag}\, \textrm{arcsec}^{-2}$ (upper
$B$-band magnitude limit in Barazza et al. 2003).
     
 \begin{table} \label{newre}
\centering
\begin{tabular}{l|c|c|c|}
\hline
Remnant & $r_{ep}$ & Remnant &$r_{ep}$\\
\hline
GAL1& 1.9& GAL8& 1.0\\
GAL2& 1.7& GAL9& 1.4\\
GAL3& 2.7& GAL10& 1.5\\
GAL4& 2.4& GAL11& 1.9\\
GAL5& 2.3& GAL12& 2.0\\
GAL6& 0.6& GAL13& 1.1\\
GAL7& 1.5&&\\

\hline
\end{tabular}
 \caption{Effective radii (in kpc) of the projected remnants calculated using
 the integrated light curves. In most of the cases they do not deviate
 considerably from the values obtained for the three dimensional galaxies. A
 significative exception is GAL13, a disk system viewed almost edge on.  }
\end{table}

The surface brightness profiles, obtained by differentiating the
cumulative light curve with respect to the equivalent radius $r_p$
(Barazza et al. 2003), are plotted in Fig. 12. 
The oscillation in the profiles of GAL3 and GAL12 reveals
the presence of spiral arms viewed almost face on (Fig. 14), which enhance
the local stellar density and produce abrupt changes in the
position angle and eccentricity profiles.
\begin{figure} \label{double}
\epsfxsize=8truecm
\epsfbox{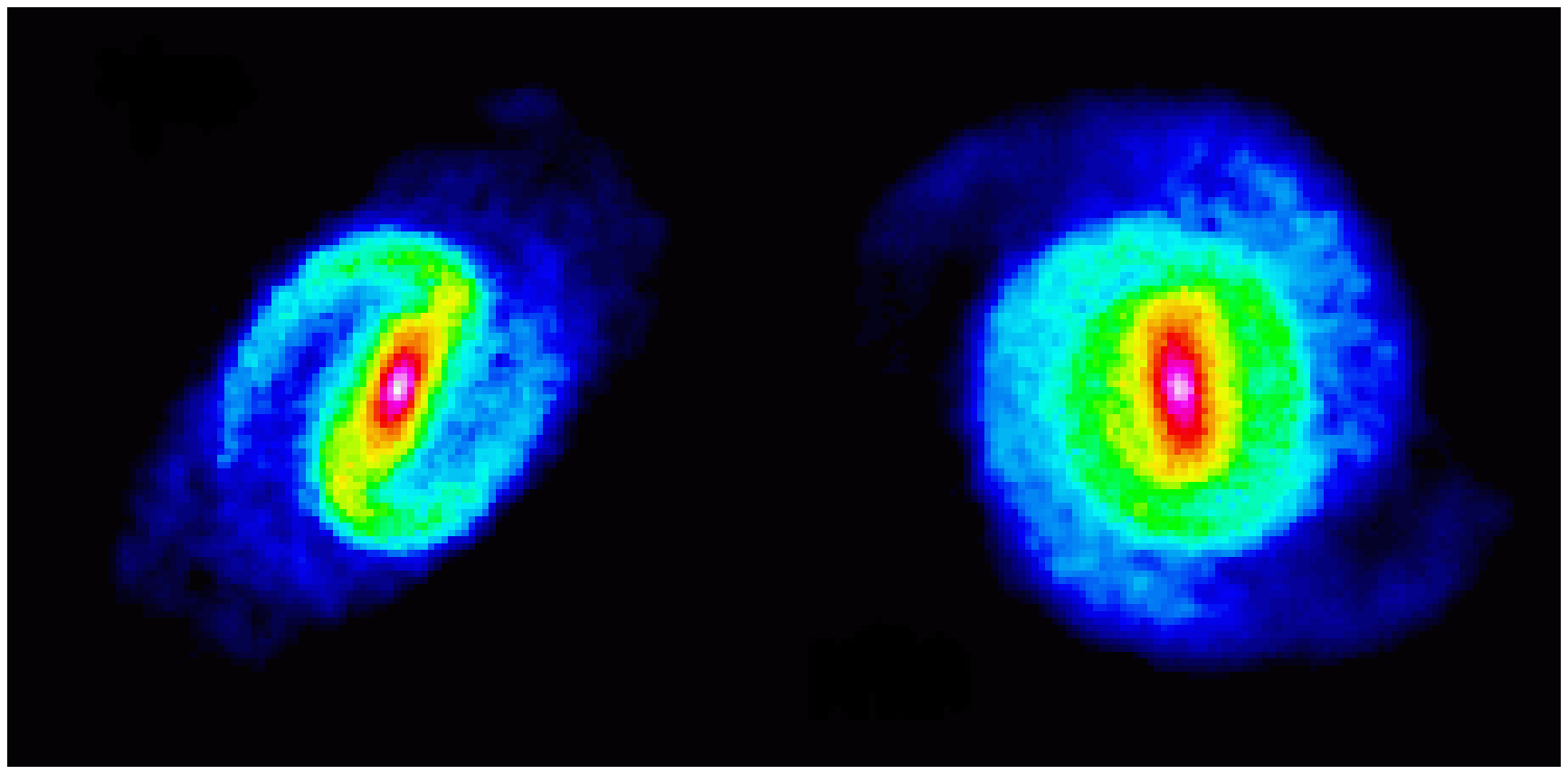}
\caption{Surface density distribution of the two remnants presenting a
  nearly face on spiral pattern (GAL3 on the left and GAL12 on the right). The physical size of both the images is $\sim 12$ kpc.  }
\end{figure}
The surface brightness profiles between 0.1 kpc (softening length for the
stellar component) and the isophotal level of $27 \, \textrm{mag} \,
\textrm{arcsec}^{-2}$ were fitted with a S\'ersic profile
\begin {equation}\label{Sersicsigma}
\sigma(r)=\sigma_0 e^{{(-r/r_0)}^{(1/n)}},
\end{equation}
where $\sigma$ is the surface brightness and $\sigma_0$ (central surface brightnes), $r_0$ (scale length) and $n$ (shape parameter) are
the free parameters of the fit. For $n = 1$ the S\'ersic model corresponds to an
exponential law, while a shape parameter $n = 4$ is indicative of a de
Vaucouleurs profile.  
Converting surface brightnesses to magnitudes, Equation \ref{Sersicsigma} becomes\begin{equation}\label{Sersicmu}
\mu(r) = \mu_0 + 1.086 (r/r_0)^{(1/n)},
\end{equation}
where $\mu_0$ is the magnitude corresponding to $\sigma_0$.
 
The fitting curves and parameters of those galaxies that it was possible to fit with a S\'ersic model are indicated in Fig. 12. 
The range of surface brightness parameters agrees quite well with the observations of dwarf galaxies in Virgo.
The central surface brightness of our remnants is on average slightly
higher than the values observed in $B$-band by Barazza et al. 2003, who fitted
S\'ersic profiles over similar radial intervals, the scale lengths obtained from
the fitting procedure are comparable with their results.    
Also note that the quoted central surface brightness values are Sercic fits
extrapolated to $r=0$. The actual values at our resolution limit are somewhat fainter (Table 3).
Galaxies with the smallest values of the shape parameter $n$ are not well
fitted by a S\'ersic profile in the central part. This is the case of GAL4 --
apparently a pure spheroidal system with high ellipticity, in reality
characterised by a disk component seen edge on -- and of GAL7, which is still 
an asymmetric bar like system with a large central nuclear component with small 
ellipticity values. For GAL7 we also show surface brightness profiles, ellipticity 
and position angles for the face on and the end on projections (Fig. 13). The smallest 
central ellipticity is associated with the end on projection, where the bar is 
viewed along the major axis. Only the less massive remnants GAL6 and GAL8
have surface brightness profiles approaching to exponentials ($n\sim1$).
   
\begin{table} 
  \begin{center}
\begin{tabular}{l|c|c|c|c}
\hline
  Galaxy& $\mu_0$  &$\mu$ (0.1 kpc) &$r_0$ &$n$   \\

 \hline
GAL1  & 19.12 & 20.50 & 0.0658 & 1.953 \\
GAL2  & 18.87 & 20.38 & 0.0478 & 2.208 \\ 
GAL4  & 16.27 & 19.44 & 0.00351 & 3.115  \\
GAL5  & 18.21 & 20.24 & 0.0181 & 2.740  \\
GAl6  & 21.37 & 21.79 & 0.284 & 0.949  \\
GAL7  & 15.58 & 18.95 & 0.00431 & 2.778  \\
GAL8  & 20.27 & 20.60 & 0.354 & 1.054  \\
GAL9  & 18.52 & 20.02 & 0.0529 & 1.972 \\
GAL10 & 18.91 & 20.39 & 0.054 & 2.008\\
GAL11 & 18.79 & 20.01 & 0.0797 & 1.957  \\
GAL13  & 20.69 & 20.83 & 0.607 & 0.885  \\ 
  
\hline
\end{tabular}
 \caption{Parameters of the S\'ersic fit: central surface brightness (in mag arcsec$^{-2}$), surface brightness at the softening radius, scale length (kpc) and shape parameter. }
  \end{center}
\end{table}

\section{Kinematics}
\begin{figure*} 
\vskip 11.2 truein
\hskip -9.0 truein
\includegraphics{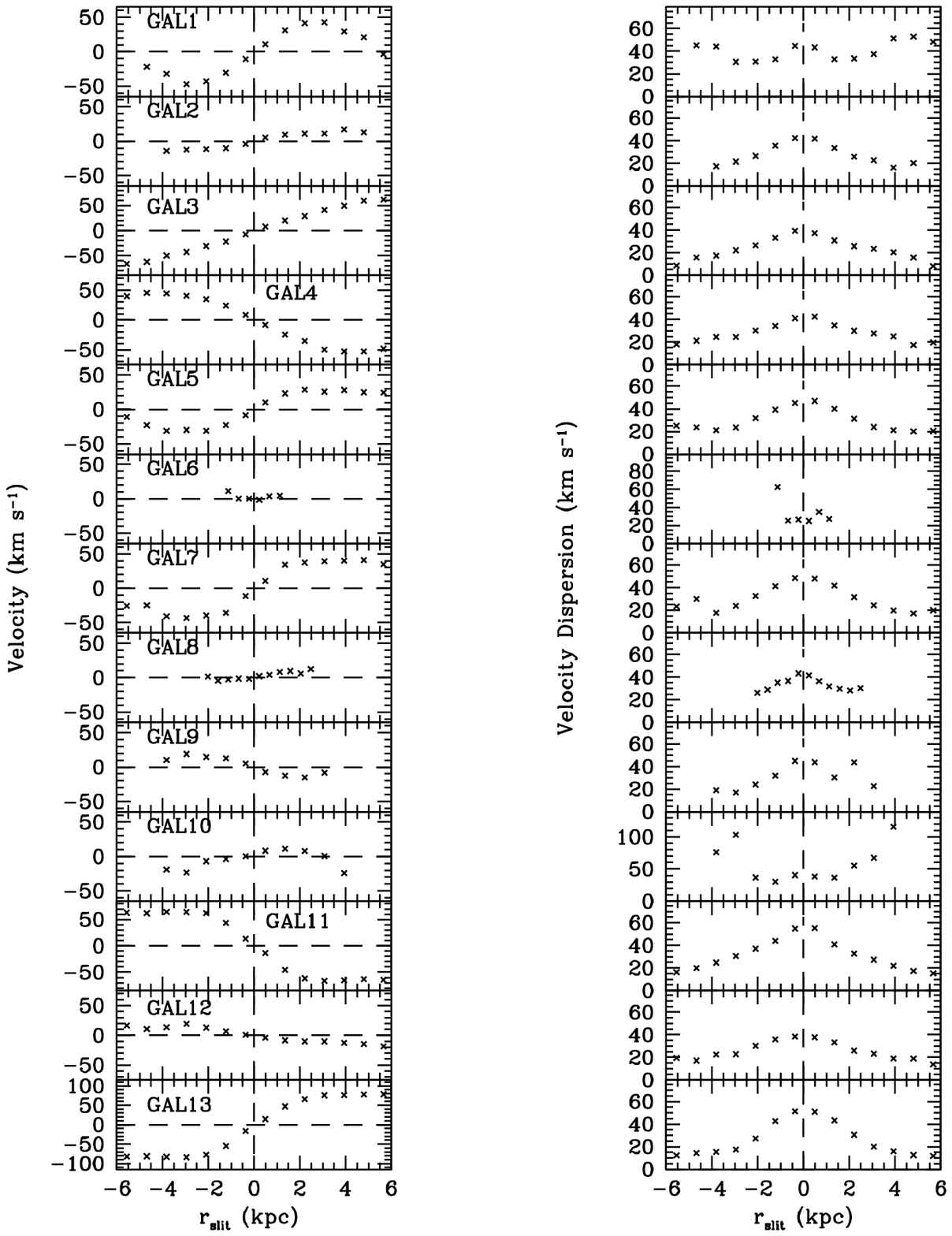}
\vspace{-6.5cm}
\caption{Projected kinematic profiles of the remnants within the cluster virial radius at $z=0$. The line of sight velocity (left panels) and the velocity dispersion (right panels) are plotted as a function of the radial distance along the major axis.}
\end{figure*}

\begin{figure*} 
\vskip 11.2 truein
\hskip -9.0 truein
\includegraphics{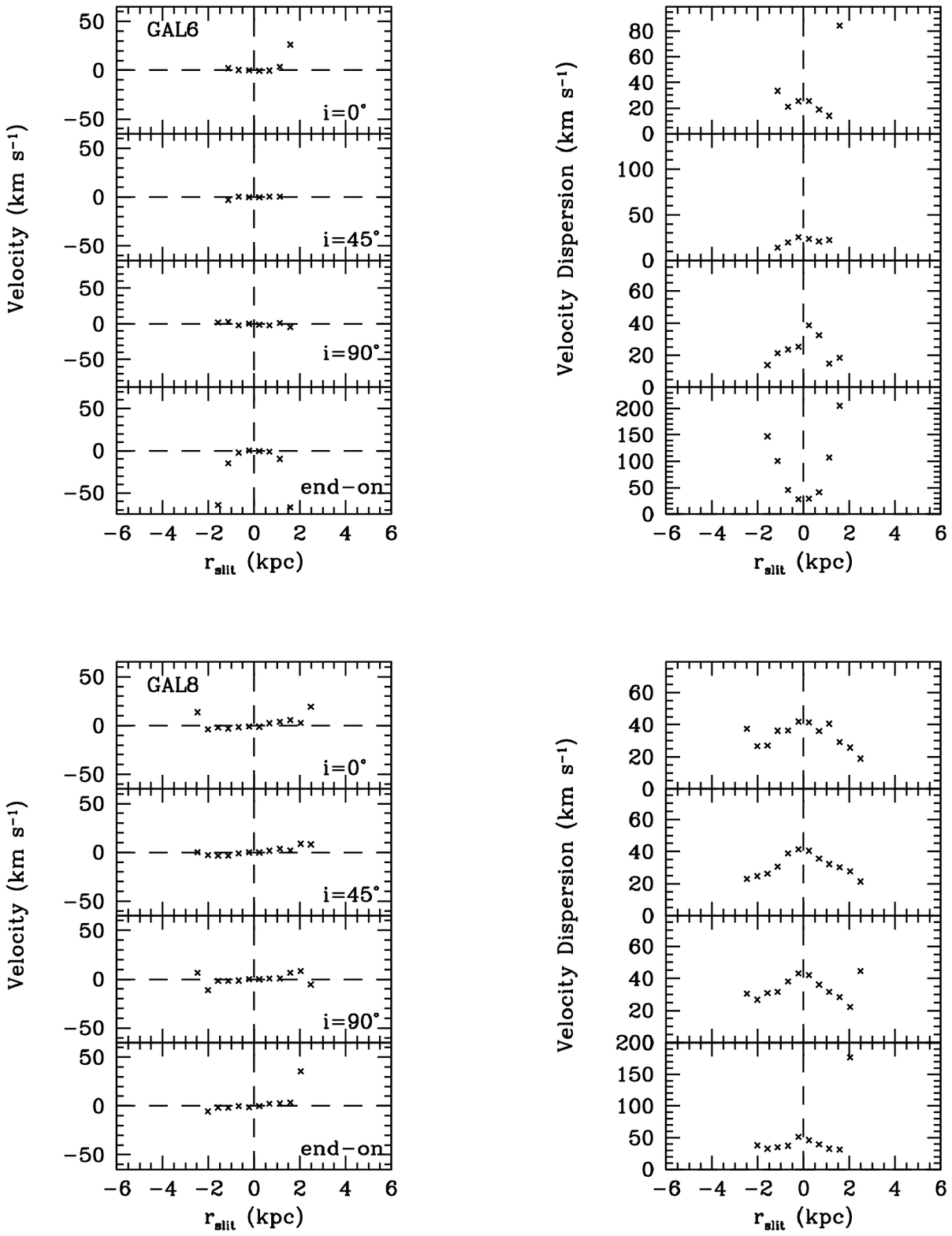}
\vspace{-6.5cm}
\caption{Projected kinematic profiles of GAL6 and GAL8 for $i=0^{\circ}$ (face-on orientation), $i=45^{\circ}$, $i=90^{\circ}$ (edge-on) and $i=90^{\circ}$ with the three dimensional major axis aligned with the line of sight (end-on projection). The line of sight velocity (left panels) and the velocity dispersion (right panels) are plotted as a function of the radial distance along the projected major axis.}
\end{figure*}

We attempt to measure the kinematics of the remnants in a similar way as previous authors measuring dwarf galaxies in the Virgo cluster.
We measure the kinematics of the 13 projected galaxies analyzed in Section 4 using a 0.3 x 12 kpc slit, which corresponds, at the distance of Virgo (15.3 Mpc,
\citet{Freedman}) to 3'' x 40''. This radial range is comparable with the slit length adopted by van Zee et al. 2004 and allows us to observe in most of the galaxies the turnover of the rotation curve, while the thickness of the slit was set to the softening length of 0.1 kpc.  
The slit was positioned along the major axis of the projected remnant centered on the galaxy and star particles within the slit area were binned into a grid. For each
grid interval with more than 15 particles -- excluding thus the most weakly bound particles at the edges of the system --  we calculated the mean particle velocity (relative to the center of mass of the galaxy) and velocity dispersion along the line of sight. For the two smallest remnants GAL6 and GAL8 we increased the number of bins in order to obtain more precise velocity curves. The resulting kinematic profiles are indicated in Fig. 15, where for each galaxy the line of sight velocity (left panels) and velocity dispersion (right panels) are plotted as a function of the radial distance along the major axis.   
In most of the cases the remnants lose a large fraction of rotation
velocity, but only for one galaxy (GAL6) 
we do not observe any significative rotation along the major axis. GAL8 shows a well defined symmetric rotation pattern although its peak
velocities is only $\sim 7$ km s$^{-1}$. GAL6 and GAL8 do not show evidences for structures or disk features and are spheroidal systems with a c/a ratio larger (Table 1) than the other remnants and surface brightness profiles well fitted by exponential laws.  Moreover, they are quite small ($M_s \leq 1.5 \times 10^9 M_{\odot}$) baryonic dominated galaxies presently orbiting at $\approx 100$ kpc from the cluster center.

Is the lack of rotation observed in GAL6 due to a projection effect, i.e. we are looking at the face on projection of a rotating system, or is it physically significative? Fig. 16 illustrates the kinematic profiles of GAL6 and GAL8 for different inclinations of the remnant: from the top to the bottom $i=0^{\circ}$, which corresponds to a face-on projection, $i=45^{\circ}$ and $i=90^{\circ}$ (edge-on). The last panel on the bottom represents an end-on galaxy, with inclination $i=90^{\circ}$ and the three dimensional major axis aligned with the line of sight. GAL6 shows traces of rotation only for $i=0$, with a maximum rotational velocity $v_{max}=3.7 \,\textrm{km}\,\textrm{s}^{-1}$, while the other projections have not well defined rotation curves and values of $v_{max} \lsim \, 2.7 \,\textrm{km}\,\textrm{s}^{-1}$. In the case of non-rotating galaxies we estimated an upper limit for $v_{max}$ by differentiating the average velocities on each side of the major axis and dividing by two \citep{Geha02}. In the calculation of $v_{max}$ for $i=0^{\circ}$ and $i=90^{\circ}$ we excluded points with velocity dispersion larger than $80 \,\textrm{km}\,\textrm{s}^{-1}$, representing stars close to the tidal radius of the remnant, which projected motion respect to the center of mass is due to stripping processes and not to rotation.  
On the other hand GAL8 has always a well defined rotation pattern and there is not a preferential direction with zero rotation.
Among the other galaxies characterised by low rotation velocities (see Fig. 15), GAL2, GAL9 and GAL10 are more massive systems if compared with the two spheroids GAL6 and GAL8, but have still high c/a values, while the low line of sight velocity of GAL12 is not related to a morphological transformation but is simply due to the fact that it is a spiral galaxy viewed face on (Fig. 14).

On the basis of this limited sample we expect non-rotating and slowly rotating
dwarfs to have internal and environmental properties substantially different
from the rotating ones. In particular the former seem to be the product of a
violent morphological transformation accelerated by the proximity of the cluster center. 
This result is apparently in contradiction with Geha et al. 2003 who finds no
relationship between the amount of rotation and the morphology or the orbital
properties of the galaxies in their sample. 

From Fig. 15 it appears evident that although the galaxies lose significant amounts of mass, the
velocity dispersion can rise up to
$50 \,\textrm{km}\,\textrm{s}^{-1}$ in the central 1-2 kpc. This is due to the 
bar formation and the transformation of circular to more radial orbits. The velocity dispersion profiles are typically decreasing at larger
radii but they rise again in those galaxies for which the slit semi-length approaches or exceeds the tidal radius. 

As the ratio between rotational velocity and velocity dispersion decreases,
the galaxy loses rotational support and the flattening becomes dominated by anisotropic pressure. The degree of rotational
support can be expressed in terms of the anisotropy parameter $(v/\sigma)^*$
\citep{Binney78}, defined as 
\begin{equation} \label{rotsupport}  
(v/\sigma)^* = \frac{v/\bar{\sigma}}{\sqrt{\epsilon(1-\epsilon)}},
\end{equation}
where $v$ is the rotational velocity, $\bar{\sigma}$ and $\epsilon$ the mean
projected velocity dispersion and ellipticity.
The theoretical prediction for a rotationally flattened
spheroidal is given by $(v/\sigma)^* \sim 1$ \citep{Binney87} and is
represented in Fig. 20 by a continuous solid line. Galaxies well
below this curve have significant velocity anisotropy.
\begin{figure} 
\epsfxsize=8truecm
\epsfbox{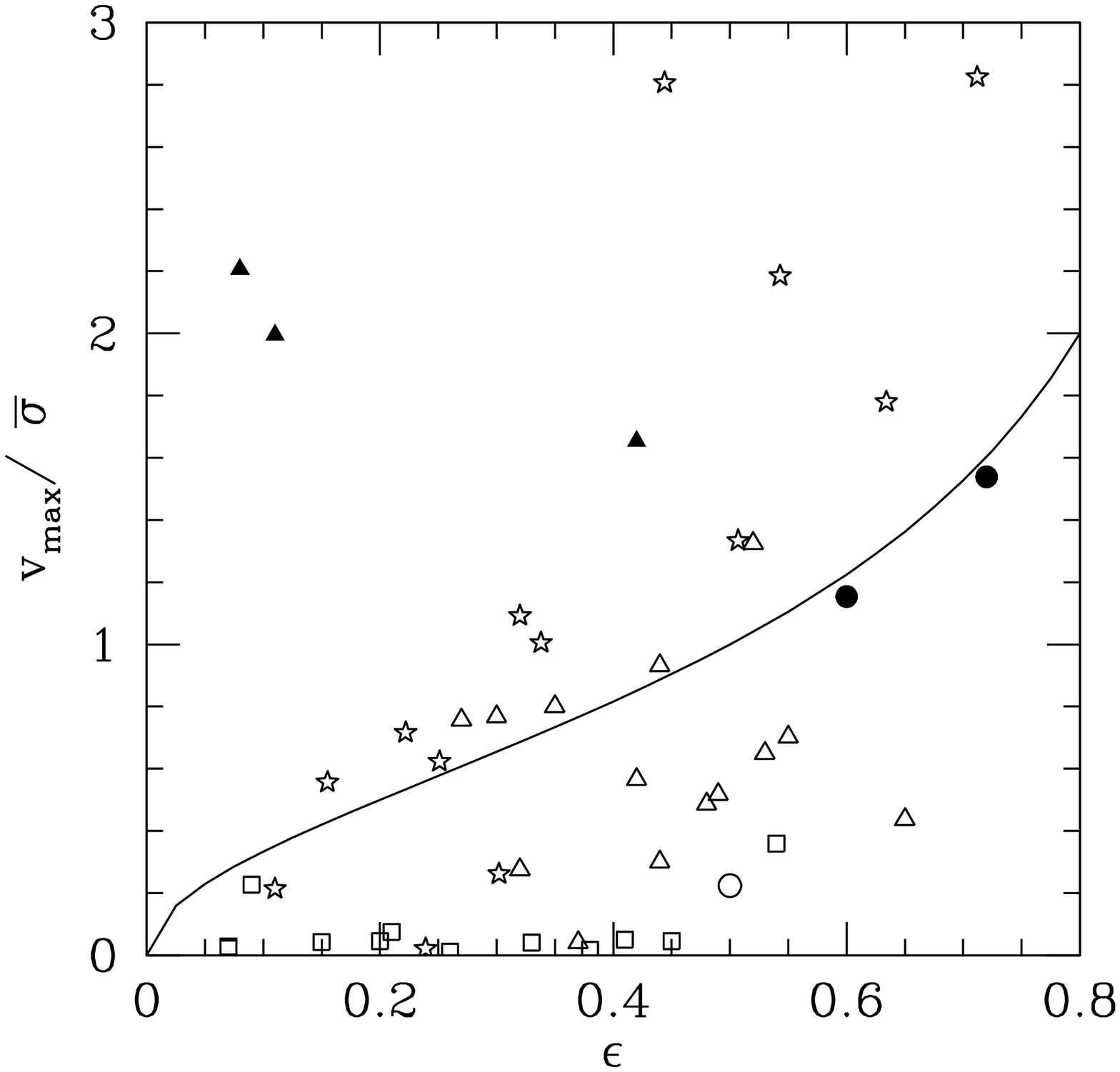}
\caption{The ratio of the maximum rotational velocity $v_{max}$ to the mean
  velocity dispersion $\bar{\sigma}$ plotted versus the mean isophotal
  ellipticity. Stars represent the simulated sample, while open triangles,
  open squares, filled triangles and open circles show observational data
  from \citet{Zee}, \citet{Geha03}, \citet{Pedraz} and \citet{Bender},
  respectively. For galaxies observed both by \citet{Geha03} and \citet{Zee} we assumed the values of $v_{max}$ and $\bar{\sigma}$ provided by the latter work. Filled circles indicate data relative to dwarf spheroidals in
  the Fornax cluster from \citet{DeRijcke03}. The solid line represents the theoretical prediction for a galaxy flattened by rotation from \citet{Binney87}. }
\end{figure}

\begin{figure} 
\epsfxsize=8truecm
\epsfbox{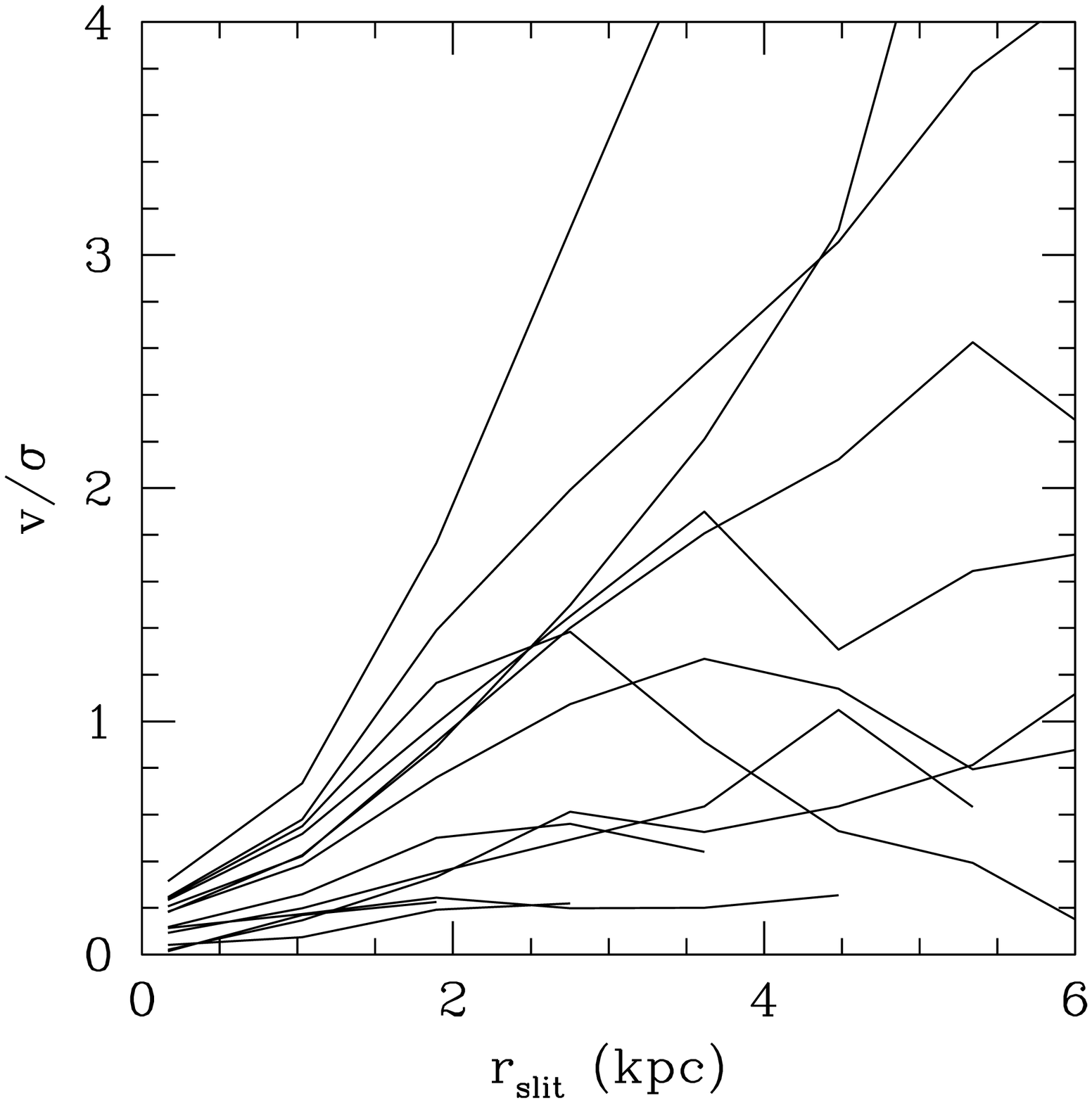}
\caption{$v/\sigma$ ratio as a function of the radial distance along the major axis for the 13 projected remnants of Fig.15. }
\end{figure}
In order to compare our results with observations, we plotted in Fig. 17 the ratio of the maximum rotational velocity $v_{max}$ to the mean
velocity dispersion $\bar{\sigma}$ within a radius of 6 kpc versus the mean
isophotal ellipticity calculated in the same radial range. In the same plot we
show the observational results for most of the dwarf galaxies in Virgo with published kinematics. We also plot the data for dwarf galaxies in the Fornax cluster by \citet{DeRijcke03}. 
Some of the galaxies in Fig. 17 (VCC 543, VCC 917, VCC 1036, VCC 1261 and VCC 1308) were observed by both Geha et al. 2003 and van Zee et. al 2003, using the Keck II  and Palomar telescope. The observations of van Zee et. al 2003, which cover a larger radial extent due to the aperture size, gives to significantly higher values of $v_{max}$ respect Geha et al. 2003. This appears to be due to the fact that the smaller slit used by the latter authors only samples the inner rising part of the rotation curve and does not include the maximum value. For example VCC 917 and VCC 1308, classified as non-rotating galaxies (both with $v_{max}=0.4$ km s$^{-1}$) by Geha et al. 2003, have $v_{max}>10$ km s$^{-1}$ according to van Zee et al. 2003, while the rotation curves of the slowly rotating dwarfs VCC 543 and VCC 1036 ($v_{max}=13$ and $14$ km s$^{-1}$ in Geha et al. 2003) peak beyond 15'' with $v_{max}>40$ km s$^{-1}$. Most of the simulated remnants have a similar range of ellipticities and $v_{max}/\bar{\sigma}$ as the observed dwarfs, even if it is clear that we have some difficulties in reproducing galaxies with a degree of rotational support close to zero. On the other hand it is possible that the low $v_{max}/\bar{\sigma}$ ratio of some galaxies is due to the small radial range covered by the observations.  
In order to understand how much the slit length influences the results, in Fig. 18 we plotted the $v/\sigma$ ratio of the projected remnants as a function of the radial distance along the slit. In most of the cases $v/\sigma$ increases with radius (with the remarkable exception of GAL1 whose rotational support drops beyond $r=2.5$ kpc). Using a 4 kpc slit we would find 10 galaxies below $v_{max}/\bar{\sigma}=1$, 6 of which with $v_{max}/\bar{\sigma}<1$ and 3 well below $v_{max}/\bar{\sigma}=0.1$, which is the upper limit assumed by Geha et al. 2003 for non-rotating dwarfs. Increasing the slit length up to 8 kpc, the number of galaxies with a low degree of rotational support is still quit high (7 galaxies with $v_{max}/\bar{\sigma}<1$ and 4 with $v_{max}/\bar{\sigma}<0.5$), but only GAL6 has still a $v_{max}/\bar{\sigma}\sim 0$, as in the case of the adopted slit length of 12 kpc.    

\section{Conclusions}

We follow the evolution of late type disk galaxies as they evolve within a
hierarchically forming galaxy cluster using N-body simulations.
We use a single galaxy model 
to explore the effects of cluster-centric position and orbit. 
Our resolution is such that we
can compare with Keck and Palomar kinematics and VLT imaging data. 
We find that:

(i) The cluster environment can be responsible for triggering bar and buckling
instabilities. 

(ii) Tidal shocks with dark matter substructures and the mean
cluster potential are both important at driving the evolution from disks
to spheroidal structures. However, galaxies within the inner 100 kpc suffer the
most complete transformation.

(iii) None of the disks are completely destroyed therefore we do not expect
to create UCD galaxies by progenitor disks with cuspy dark halos \citep{Bekki}.

(iv) More tidal heating leads to more mass loss which leads to rounder galaxies.

(v) The kinematics ($v/\sigma$) of the stellar remnants is similar to observations
of Virgo and Fornax dE/dSph galaxies. Only the most severely disrupted galaxies
which lose most of their stellar mass become completely pressure supported. Most
of the remnants retain a significant amount of rotational motion.

(vi) Unsharp masked images of the galaxies show similar hidden bar and spiral
features as some cluster galaxies. These gradually dissapear as the
galaxies suffer more heating.
Our results could be rescaled in time and mass to make statements about possible
disk progenitors entering group or even galactic environments. The next step is
to self-consistently follow the evolution of galaxies as they form and evolve
in different environments.
 
Our study has focussed on the evolution of a late type
disk similar to the Large Magellanic Cloud. The baryon
fraction in the disk is relatively high and adiabatic contraction
has enabled the galaxy to be more robust. Disks such as M32, have
a baryon fraction about five times lower and thus will be more
susceptible to tidal perturbations. Exploring a large parameter
space is beyond the scope of this paper, however we note that
if we followed the evolution of fainter, baryon poor disks
we would expect a more complete transformation to dE/dSph and a
lower fraction of galaxies with hidden features such as bars
and disks.

\section{Acknowledgements}
We would like to thank J\"urg Diemand for providing the cosmological cluster simulation. We are grateful to Fabio Barazza, Marla Geha, Frank van den Bosch, Stelios Kazantzidis and Andrea Macci\'o for useful discussions. The numerical simulations were performed on the Zbox (http:/www.theorie.physik.unizh.ch/stadel) supercomputer at the University of Z\"urich. CM and RP are supported by the Swiss National Science Foundation. VPD is supported by a Brooks Fellowship at the University of Washington.

\label{lastpage}


\begin{thebibliography}{99}

\bibitem[\protect\citeauthoryear{Barazza, Binggeli \& Jerjen}{2002}]{Barazza02} Barazza, F. D., Binggeli, B., Jerjen, H. 2002, A\&A, 391, 823

\bibitem[\protect\citeauthoryear{Barazza, Binggeli \& Jerjen}{2003}]{Barazza03} Barazza, F. D., Binggeli, B., Jerjen, H. 2003, A\&Ap, 407, 121 

\bibitem[\protect\citeauthoryear{Bekki et al.}{2003}]{Bekki} 
Bekki, K., Couch, W. J., Drinkwater, M. J., Shioya, Y. 2003, MNRAS, 344, 399 



\bibitem[\protect\citeauthoryear{Bender \& Nieto}{1990}]{Bender} 
Bender, R., Nieto, J. L. 1990, A\&A, 239, 97 



\bibitem[\protect\citeauthoryear{Binney}{1978}]{Binney78} 
Binney, J. 1978, MNRAS, 183, 501 


\bibitem[\protect\citeauthoryear{Binney \& Tremaine}{1987}]{Binney87} 
Binney, J., Tremaine, S. 1987, Princeton, NJ, Princeton University Press, 1987  

\bibitem[\protect\citeauthoryear{Conselice, Gallagher \&
    Wyse}{2001}]{Conselice}
Conselice, C. J., Gallagher, J. S., Wyse, R. F. G. 2001, ApJ, 559, 791 

\bibitem[\protect\citeauthoryear{Courteau}{1997}]{Courteau} 
Courteau, S. 1997, AJ, 114, 2402 

\bibitem[\protect\citeauthoryear{De Rijcke et al.}{2001}]{DeRijcke01} 
De Rijcke, S., Dejonghe, H., Zeilinger, W. W., Hau, G. K. T. 2001, ApJ, 559, L21 

\bibitem[\protect\citeauthoryear{De Rijcke et al.}{2003}]{DeRijcke03} 
De Rijcke, S., Dejonghe, H., Zeilinger, W. W., Hau, G. K. T. 2003, A\&A, 400, 119 


\bibitem[\protect\citeauthoryear{Diemand, Moore \& Stadel}{2004a}]{Diemand04a}
Diemand, J., Moore, B., Stadel, J. 2004a, MNRAS submitted (astroph 0402267)

\bibitem[\protect\citeauthoryear{Diemand, Moore \& Stadel}{2004b}]{Diemand04b}
Diemand, J., Moore, B., Stadel, J. 2004b, MNRAS, tmp. 152D


\bibitem[\protect\citeauthoryear{Dressler et al.}{1994}]{Dressler} 
Dressler, A., Oemler, A. J., Butcher, H. R., Gunn, J. E. 1994, ApJ, 430, 107 

\bibitem[\protect\citeauthoryear{Ferguson \& Sandage}{1991}]{Ferguson91} 
Ferguson, H. C., Sandage, A. 1991, AJ, 101, 765 

\bibitem[\protect\citeauthoryear{Freedman et al.}{2001}]{Freedman} 
Freedman, W. L., \emph{et al.} 2001, ApJ, 553, 47 

\bibitem[\protect\citeauthoryear{Gavazzi et al.}{2004}]{Gavazzi}
Gavazzi, G. \emph{et al.}, astroph 0410228
Accepted to A\&A

\bibitem[\protect\citeauthoryear{Geha, Guhathakurta \& van der Marel}{2002}]{Geha02} 
Geha, M., Guhathakurta, P., van der Marel, R. P. 2002, AJ, 124, 3073

\bibitem[\protect\citeauthoryear{Geha, Guhathakurta \& van der Marel}{2003}]{Geha03} 
Geha, M., Guhathakurta, P., van der Marel, R. P. 2003, AJ, 126, 1794 

\bibitem[\protect\citeauthoryear{Gnedin}{2003}]{Gnedin} 
Gnedin, O.Y. 2003, ApJ, 589, 752 



\bibitem[\protect\citeauthoryear{Graham, Jerjen \& Guzm{\' a}n}{2003}]{Graham} 
Graham, A. W., Jerjen, H., Guzm{\' a}n, R. 2003, AJ, 126, 1787

\bibitem[\protect\citeauthoryear{Hernquist}{1993}]{Hernquist} 
Hernquist, L. 1993, ApJ, 86, 389

\bibitem[\protect\citeauthoryear{Jerjen, Kalnajs \& Binggeli}{2000}]{Jerjen} 
Jerjen, H., Kalnajs, A. \& Binggeli, B. 2000, A\&A, 358, 845 

\bibitem[\protect\citeauthoryear{Kormendy \& Freeman}{2004}]{Kormendy} 
Kormendy, J., Freeman, K. C. 2004, IAU Symposium, 220, 377 

\bibitem[\protect\citeauthoryear{Mayer et al.} {2001}]{Mayer01} 
Mayer, L., Governato, F., Colpi, M., Moore, B., Quinn, T., Wadsley, J.,
  Stadel, J., Lake, G. 2001, ApJ, 559, 754

\bibitem[\protect\citeauthoryear{Mayer et al.} {2002}]{Mayer02}
Mayer, L., Moore, F., Quinn. T., Governato, F., Stadel, J., 2002, MNRAS, 336, 119
\bibitem[\protect\citeauthoryear{Mo, Mao, \& White}{1998}]{Mo} 
Mo, H. J., Mao, S., White, S. D. M. 1998, MNRAS, 295, 319 

\bibitem[\protect\citeauthoryear{Moore et al.}{1996}]{Moore96} 
Moore, B., Katz, N., Lake, G., Dressler, A., Oemler, A. 1996, Nature, 379, 613 

\bibitem[\protect\citeauthoryear{Moore, Lake \& Katz}{1998}]{Moore98} 
Moore, B., Lake, G. \& Katz, N. 1998, ApJ, 495, 139

\bibitem[\protect\citeauthoryear{Moore, Lake, Quinn \& Stadel}{1999}]{Moore99} 
Moore, B., Lake, G., Quinn, T. \& Stadel, J. 1999, MNRAS, 304, 465 

\bibitem[\protect\citeauthoryear{Moore, Diemand \& Stadel}{2004}]{Moore04}
Moore, B., Diemand, J., Stadel, J. 2004, IAU Colloquium: Outskirts of Galaxy
Clusters: intense life in the suburbs (astroph 0406615)

\bibitem[\protect\citeauthoryear{Navarro, Frenk \& White}{1996}]{Navarro96} 
Navarro J. F., Frenk C. S., White S. D. M.  1996, ApJ, 462, 563 

\bibitem[\protect\citeauthoryear{Navarro, Frenk \& White}{1997}]{Navarro97} 
Navarro J. F., Frenk C. S., White S. D. M. 1997, ApJ, 490, 493 

\bibitem[\protect\citeauthoryear{O'Neil, Bothun \& Impey}{1999}]{O'Neil} 
O'Neil, K., Bothun, G. D., Impey, C. D. 1999, AJ, 118, 1618 

\bibitem[\protect\citeauthoryear{Pedraz et al.}{2002}]{Pedraz} 
Pedraz, S., Gorgas, J., Cardiel, N., S{\' a}nchez-Bl{\' a}zquez, P., Guzm{\'
  a}n, R. 2002, MNRAS, 332, L59 
 

\bibitem[\protect\citeauthoryear{Persic \& Salucci}{1997}]{Persic} 
Persic, M., Salucci, P. 1997, ASP Conf. Ser. 117: Dark and Visible Matter in Galaxies and Cosmological Implications,  



\bibitem[\protect\citeauthoryear{Simien \& Prugniel}{2002}]{Simien} 
Simien, F., Prugniel, P. 2002, A\&A, 384, 371

\bibitem[\protect\citeauthoryear{Sparke \& Sellwood}{1987}]{Sparke} 
Sparke, L. S., Sellwood, J. A. 1987, MNRAS, 225, 653 


\bibitem[\protect\citeauthoryear{Spergel et al.}{2003}]{Spergel}
Spergel, D. N. \emph{et al.}, 2003, ApJSS, 148, 175 

\bibitem[\protect\citeauthoryear{Springel \& White}{1999}]{Springel} 
Springel, V., White, S. D. M. 1999, MNRAS, 307, 162 

\bibitem[\protect\citeauthoryear{Stadel}{2001}]{Stadel}
Stadel, J. 2001, Ph.D. Thesis

\bibitem[\protect\citeauthoryear{Toomre}{1964}]{Toomre} 
Toomre, A. 1964, ApJ, 139, 1217 


\bibitem[\protect\citeauthoryear{Toomre}{1981}]{Toomre81} 
Toomre, A. 1981, Structure and Evolution of Normal Galaxies, S. M. Fall \& D. Lynden--Bell, Cambridge Univ. Press, 111 


\bibitem[\protect\citeauthoryear{van Zee, Skillman \& Haynes}{2004}]{Zee} 
van Zee, L., Skillman, E. D., \& Haynes, M. P. 2004, AJ, 128, 121 

\bibitem[\protect\citeauthoryear{van Zee, Barton \& Skillman}{2004}]{Zee2}
van Zee, L., Barton, E. J., Skillman, E. D., astroph 0409346
Accepted to AJ.

\end{thebibliography}
\end{document}